\documentclass[journal,10pt]{IEEEtran}
\normalsize
\usepackage{gensymb}
\usepackage{lineno}
\usepackage{fancyhdr}
\usepackage{amsmath, amsthm, amssymb}
\usepackage{amsfonts}
\usepackage[dvips]{graphicx}
\usepackage[normalem]{ulem}
\usepackage{balance}
\usepackage{epsfig}
\usepackage{color}
\usepackage{bm}
\usepackage{enumerate}
\usepackage{latexsym}
\usepackage{graphics}
\usepackage{graphicx}
\usepackage{subfigure}
\usepackage{cite}
\usepackage{mathrsfs}
\usepackage{stfloats}
\usepackage{extarrows}
\usepackage{upgreek}
\usepackage{lineno}
\usepackage{caption}
\usepackage{multirow}
\usepackage[linesnumbered,ruled,vlined]{algorithm2e}
\usepackage{algpseudocode}

\newtheorem{theorem}{Theorem}

\newtheorem{corollary}{Corollary}

\newtheorem{proposition}{Proposition}

\newcommand{\vect}[1]{\mathbf{#1}}

\def\diag{\mathrm{diag}}

\def\tr{\mathrm{tr}}

\def\Htran{\mbox{\tiny $\mathrm{H}$}}
\def\Ttran{\mbox{\tiny $\mathrm{T}$}}

\begin{document}
\title{Scalable Cell-Free Massive MIMO Systems with Finite Resolution ADCs/DACs over Spatially Correlated Rician Fading Channels}
\author{
Xiangjun Ma, Xianfu~Lei, P. Takis Mathiopoulos, Kai Yu, Xiaohu Tang
\thanks{X. Ma, X. Lei, and X. Tang are with the School of Information Science and Technology, Southwest Jiaotong University, 610000 Chengdu, China (e-mails: ezio1390265506@hotmail.com, xflei81@gmail.com, xhutang@swjtu.edu.cn).}
\thanks{P. T. Mathiopoulos is with the Department of Informatics and
Telecommunications, National and Kapodistrian University of Athens,
15784 Athens, Greece (e-mail: mathio@di.uoa.gr).}
\thanks{K. Yu is with the China Railway Eryuan Engineering Group Co. Ltd, 610031 Chengdu, China
(e-mail: ekyukai@qq.com).}
}
\maketitle
\vspace{-1cm}
\begin{abstract}
In this paper, an analytical framework for evaluating the performance of scalable cell-free massive MIMO (SCF-mMIMO) systems in which all user equipments (UEs) and access points (APs) employ finite resolution digital-to-analog converters (DACs) and analog-to-digital converters (ADCs) and operates under correlated Rician fading, is presented.
By using maximal-ratio combining (MRC) detection, generic expressions for the uplink (UL) spectral efficiency (SE) for both distributed and centralized schemes are derived.
In order to further reduce the computational complexity (CC) of the original local partial MMSE (LP-MMSE) and partial MMSE (P-MMSE) detectors, two novel scalable low complexity MMSE detectors are proposed for distributed and centralized schemes respectively, which achieves very similar SE performance.
Furthermore, for the distributed scheme a novel partial large-scale fading decoding (P-LSFD) weighting vector is introduced and its analytical SE performance is very similar to the performance of an equivalent unscalable LSFD vector.
Finally, a scalable algorithm jointly consisting of AP cluster formation, pilot assignment, and power control is proposed, which outperforms the conventional random pilot assignment and user-group based pilot assignment policies and, contrary to an equal power transmit strategy, it guarantees quality of service (QoS) fairness for all accessing UEs.
\end{abstract}


\section{Introduction} \label{sec-intro}

Cell-free massive multiple-input multiple-output (CF-mMIMO) systems deploy many geographically distributed access points (APs) connected to a central procession unit (CPU) through individual fronthaul links \cite{Emil2020making}. In order to meet the demanding ubiquitous connectivity requirements set by the 6G and beyond design guidelines, they jointly serve numerous user equipment (UE) employing single time-frequency resource transmission blocks while ensuring uniform quality of service (QoS) regardless of their geographic locations \cite{Emil2020making,Ngo2017vs,Interdonato2019scalability}.
For these CF-mMIMO systems two main cooperative transmission methods have been employed for realizing the communication between the APs and CPU \cite{Emil2020making}. The first one is based upon a centralized approach where each AP acts like a relay forwarding the received pilot and data signal to the CPU which, in turn, performs in a centralized manner the channel estimation (CE) and data detection.
The second one is a distributed approach where each AP first carries out CE and data detection locally and then the detected data are sent to the CPU for further and final decoding.

In canonical CF-mMIMO sytems \cite{Emil2020making,Ngo2017vs}, there is just one CPU which connects to all APs that serve their accessing UEs. The disadvantage of such canonical systems is that they could be unscalable since the fronthaul payload is transmitted from each AP to the only one CPU. Furthermore, the computational complexity (CC) required to calculate the detection vector grows linearly as the number of the served UEs increases \cite{Emil2020scalable}.
Because of these limitations, in \cite{Interdonato2019scalability} and \cite{Emil2020scalable} a SCF-mMIMO network where all APs were divided into multiple mutually disjoint clusters, has been proposed. Each of these clusters connects to a specific CPU which is further connected to the other CPUs belonging to the same network. The main advantage of this network topology is that it can allocate the original CPU's tasks to multiple local CPUs. In this way the CC for signal processing of each AP and their fronthaul payload transmitted to their own CPU is significantly reduced as the number of UEs increases arbitrarily.
In fact, it was shown in \cite{Buzzi2020User} that, as compared to the canonical CF-mMIMO, the fronthaul payload transmitted from each AP to CPU for SCF-mMIMO can be greatly reduced.

However, as the number of APs and their equipped antennas increase, the use of high-resolution ADCs/DACs in such CF-mMIMO becomes problematic because of their excessive power consumption and hardware cost \cite{Jiayi2020cell-free-adc}. As well known, it turns out that their power consumption decreases exponentially as the resolution of ADCs/DACs decreases linearly, while the spectral efficiency (SE) loss remains low \cite{Xiangjun2021}.
Simulation results in \cite{Jacobsson2017Throughput} and \cite{Dong2017} have shown that ADCs with resolution of  only a few bits are needed to achieve SE performance comparable to the infinite-resolution case.
Motivated by these observations, we will analyze the impact of finite-resolution ADCs/DACs on the performance of SCF-mMIMO systems by presenting an analytical framework for the performance evaluation of such systems. Next, previously published papers which are related to our research topic will be briefly reviewed.

\subsection{Related Works}
\textit{\uppercase\expandafter{\romannumeral1}) Canonical CF-mMIMO:} The centralized and distributed schemes for canonical CF-mMIMO systems were first proposed in \cite{Emil2020making}, where the uplink (UL) SE performances with different detection schemes under correlated Rayleigh fading have been compared.
The closed-form expressions for the UL-SE and downlink (DL) SE under uncorrelated Rayleigh fading have been derived in \cite{Ngo2017vs} by considering greedy pilot assignment to mitigate the pilot contamination among the UEs sharing the same pilot.
Furthermore, the max-min power control was used to ensure uniform QoS delivery to all UE. In \cite{Polegre2020channel hardening}, the UL-SE for centralized scheme employing MRC detection and operating in correlated Rician fading was derived by using the use-and-then forget (UatF) bound \cite{Emil2017mMnetworks}.
However, as it does not take advantage of the estimated channel available at the CPU for signal detection, this bound is rather loose leading to inaccurate SE performance results for the centralized scheme. For this reason, the bound presented in \cite[Theorem 4.1]{Emil2017mMnetworks} will be used for our SE performance evaluation of the centralized scheme so that more accurate performance analysis can be obtained.

\textit{\uppercase\expandafter{\romannumeral2}) SCF-mMIMO:} In \cite{Emil2020scalable}, a partial minimum mean square error MMSE (P-MMSE) and a local partial MMSE (LP-MMSE) detectors operating under correlated Rayleigh fading channels have been proposed for SCF-mMIMO. Since these two MMSE detectors have been designed by only partially considering the CSIs of the accessing UEs, their CC is independent of the total number of UEs.
Furthermore, a novel scalable algorithm which jointly considers AP selection and pilot assignment was proposed. However, the equal power transmission policy proposed in \cite{Emil2020scalable} inevitably causes excessive interference to cell-edge UEs,
which makes it difficult to guarantee the QoS fairness for all UE.
In \cite{Chen2020Structured}, a SCF-mMIMO system operating in correlated Rayleigh fading has been studied and a novel low complexity partial LSFD (P-LSFD) vector achieving comparable SE performance with that of the unscalable LSFD vector in \cite{Emil2020scalable} has been proposed.
Furthermore, two pilot assignment algorithms and a fractional power control policy were introduced to suppress the mutual interference caused by the UEs sharing the same pilot and balance QoS fairness and average SE.
In \cite{Guenach2021Joint}, an iterative power control and AP selection algorithm was proposed for SCF-mMIMO to achieve QoS fairness among each UE. However, its algorithm derivation was based on the assumption that the pilot length is larger than the number of UEs and as such it cannot be effectively used for high connection scenarios such as those proposed in 6G and beyond systems.

\textit{\uppercase\expandafter{\romannumeral3}) Finite resolution ADCs/DACs:} In \cite{Jiayi2020cell-free-adc}, the UL-SE performance of CF-mMIMO with finite resolution ADCs at each AP was studied. An UL-SE expression using MRC detection was derived and a power control method which maximizes the sum SE was also presented. However, the applicability of this work is limited to distributed scheme operating under correlated Rayleigh fading. Furthermore, in its analysis the scalable scenario was not considered.
In \cite{Hu2019Cell}, the DL-SE performance of a CF-mMIMO using MRT precoding and APs and UEs employing finite-resolution ADCs has been derived.  Furthermore, a power control scheme which maximizes the minimum DL-SE of all UEs was also proposed to improve the QoS fairness. In \cite{Zhang2019on the performance}, a UL-SE expression for CF-mMIMO using MRC detector with APs employing finite-resolution ADCs was derived, and the tradeoff between UL-SE and UL-EE was investigated. It is noted that since \cite{Hu2019Cell} and \cite{Zhang2019on the performance} have mainly focused on the SE analysis for distributed schemes operating under independent Rayleigh fading, and analysis is limited to canonical CF-mMIMOs.

\subsection{Motivation and Contributions}
From the above literature review on the topic of CF-mMIMO systems,
it is clear that the channel between UE and AP has been often modelled as a correlated Rayleigh fading (e.g., see \cite{Emil2020making}, \cite{Emil2020scalable}, and \cite{Chen2020Structured}). However, as in practice the antennas exhibit nonuniform radiation patterns, their gain is not uniformly distributed in all channel directions \cite[Definition 2.3]{Emil2017mMnetworks}, and thus the Rayleigh fading channel model might not be realistic. Furthermore, since the trend is to increase the density of the AP clusters, it is more appropriate to consider instead a Rician fading channel as it is very likely that there will be an additional direct line-of-sight (LOS) link between the UE and the APs \cite{Ozdogan2019Massive,Polegre2020channel hardening}.
More importantly, although both the P-MMSE and LP-MMSE detectors in \cite{Emil2020scalable} and \cite{Chen2020Structured} are designed irrelevant to the total number of accessing UEs, their required CCs are still large due to the excessive CE costs.

Motivated by the discussion above, in this paper, we analyze and evaluate the SE performance of a SCF-mMIMO under correlated Rician.
Besides, we also consider a more realistic case where each AP and UE is equipped with finite resolution ADCs/DACs. In the past only ideal ADCs/DACs over correlated Rayleigh fading have been considered for the performance analysis of SCF-mMIMO systems.
Our approach will be based on the development of an analytical framework which, apart from providing analytical performance evaluation results, will make the following specific contributions: \textit{i}) Provide effective cluster formation scheme to design the novel scalable LSFD weighting vector and the low complexity scalable MMSE detectors whose SE performances are very close to the previous scalable MMSE detectors; \textit{ii}) Propose effective pilot assignment, and power control strategy to simultaneously suppress the pilot contamination and guarantee the QoS fairness; and \textit{iii}) Investigate the impact of non-ideal ADCs/DACs on the SE performance.
The detailed contributions can be summarized as follows.
\begin{itemize}
\item
We provide an analytical framework for evaluating the performance of a SCF-mMIMO with finite resolution ADCs/DACs for both distributed and centralized schemes. Specifically, using MRC detection and considering operation in correlated Rician fading, generic UL-SE expressions valid for arbitrary AP cluster formation, pilot assignment, and transmit power will be derived. Performance evaluation results will show that the SE performance improvement is limited for ADCs/DACs resolution higher than $4$-bits.
Furthermore it will be shown that the centralized scheme yields higher SE only under Rayleigh fading, while the distributed scheme yields higher SE under Rician fading.
\item
For the proposed SCF-mMIMO system, two low complexity LP-MMSE and the P-MMSE detectors for distributed and centralized schemes are proposed.
It will be shown that the proposed detectors, as compared to previously proposed scalable MMSE detectors, not only have lower CC but also yield similar SE performances.
In addition, based on the derived SE expression using MRC for the distributed scheme, a novel scalable P-LSFD weighting vector yielding very similar SE performance, as compared to the unscalable LSFD vector, is proposed.
\item
By jointly considering AP cluster formation, pilot assignment, and UL power control for each UE, a novel scalable algorithm is proposed in order to suppress the pilot contamination caused by the UEs sharing the same pilot and simultaneously achieve the desirable tradeoff between the QoS fairness and average SE. Simulation results will show that, on one hand, the proposed pilot assignment scheme yields much higher SE performance as compared with the random pilot assignment and user-group based pilot assignment strategies. On the other hand, the QoS fairness of each UE can be greatly achieved in a small amount of average SE performance loss as compared to the traditional equal power transmit strategy.
\end{itemize}

\subsection{Paper Outline and Notation}
Our paper is organized as follows. After this introduction, the system model for our proposed system is presented in Section \ref{section:System Model}. Then, the scalable detection methods both for distributed and centralized connection scheme are introduced and the closed form of UL-SE expressions using MRC detection are derived in Section \ref{section:Spectral Efficiency Analysis}. In Section \ref{section:algorithm}, the joint algorithm for AP cluster formation, pilot assignment, and UL power control, is proposed. Furthermore, in Section \ref{section:simulation}, numerous performance simulation results are provided to corroborate our analytical results and give more insights into the UL-SE performance under different channel conditions, pilot assignment and power control strategies. Finally, conclusions are given in Section \ref{section:conclusion}.

\subsubsection*{Notation}
The matrix and vector are denoted by upper and lower case boldface, respectively. $[\vect{X}]_{i_1i_2}$ denotes the element at $i_1^\mathrm{th}$ row and $i_2^\mathrm{th}$ column of matrix $\vect{X}$. The Frobenius and $2$ norms of matrix $\vect{X}$ are represented by $\| \vect{X} \|_\mathrm{F}$ and $\| \vect{X} \|$, respectively.
$\tr(\vect{X})$ represents the trace of matrix $\vect{X}$. The $N \times N$ identity matrix is denoted by $\vect{I}_N$. The diagonal matrix of $\vect{X}$ is denoted by $\diag(\vect{X})$.
$\left[r_{i}\right]_{K}^\Lambda$ is the $K$-order diagonal matrix whose $i^\mathrm{th}$ diagonal element is $r_{i}$, $\left[r_{i_1i_2}\right]_{K}^\mathrm{S}$ is the $K$-order square matrix whose element at $i_1^\mathrm{th}$ row and $i_2^\mathrm{th}$ column is $r_{i_1i_2}$, and $\left[\vect{R}_i\right]_{K}^\mathrm{B}$ is $K$-order block-diagonal matrix whose $i^\mathrm{th}$ block element is $\vect{R}_i$.
$\mathcal{CN}(\vect{0},\vect{R})$ represents the circularly symmetric complex Gaussian distribution with zero mean and covariance matrix $\vect{R}$. The superscripts $^{\Ttran}$, $^\ast$, $^{\Htran}$, and $\dagger$ denote transpose, conjugate, Hermitian transpose, and pseudoinverse, respectively. $\Re\{\cdot\}$ presents the real part of a complex number.

\section{System Model}\label{section:System Model}
\begin{figure}
\begin{center}
\includegraphics[width=0.8\columnwidth]{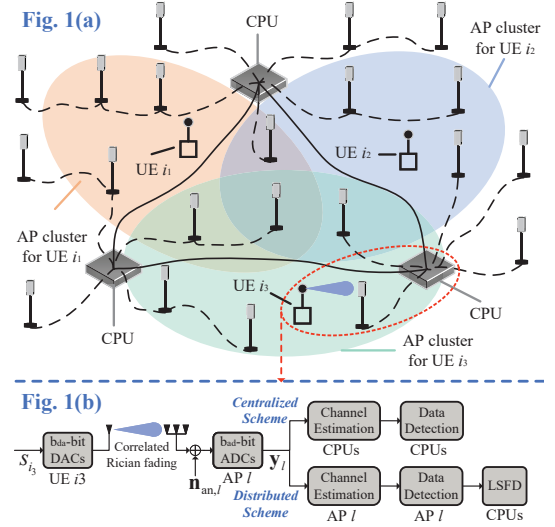} 
\captionsetup{font={normal}}
\caption{The system model of the SCF-mMIMO under consideration assuming three arbitrary AP clusters. (a) Cluster formation. (b) Transceiver structure model for the communication link between UEs, APs, CPUs.}\label{fig_system_model}
\end{center} \vskip-7.7mm
\end{figure}

As illustrated in Fig. \ref{fig_system_model}, the considered TDD SCF-mMIMO consisting of $K$ uniformly distributed single-antenna UEs and $L$ APs each equipped with $N$ antennas is presented. All $L$ APs are divided into disjoint sets and each set of APs connects to a common CPU via fronthaul links where all CPUs are interconnected with each other \cite{Emil2020making}.
In order to ensure that the CC at each AP and CPU does not growing arbitrarily large as $K\rightarrow\infty$, each UE is served by a particular cluster of APs.
In Fig. \ref{fig_system_model}(a), each of the three colored region contains a specific cluster of APs which cooperatively serve one particular UE co-located in this area. Such operation is termed as user-centric cluster formation \cite{Chen2020Structured} and the detailed AP cluster formation scheme will be discussed in Section \ref{section:algorithm}.
In Fig. \ref{fig_system_model}(b), the transceiver structure for the communication links between the UEs, APs and CPUs is presented. The correlated Rician fading channel between $l^\mathrm{th}$ AP and $k^\mathrm{th}$ UE, $\vect{h}_{kl}=\vect{\bar{h}}_{kl}+\vect{h}_{\mathrm{w},kl}\in\mathbb{C}^N\sim\mathcal{CN}(\vect{\bar{h}}_{kl},\vect{R}_{kl})$ can be decomposed into a deterministic LOS part $\vect{\bar{h}}_{kl}$ and a random non-line-of-sight (NLOS) part $\vect{h}_{\mathrm{w},kl}$ where $\vect{h}_{\mathrm{w},kl}\sim\mathcal{CN}(\vect{0},\vect{R}_{kl})$ remains constant within $\tau_c$ symbols. Note that $\vect{R}_{kl}=\mathbb{E}[\vect{h}_{\mathrm{w},kl}\vect{h}_{\mathrm{w},kl}^\mathrm{H}]\in\mathbb{C}^{N\times N}$ denoting the positive semi-definite spatial correlation matrix which is assumed to be locally known at $l^\mathrm{th}$ AP \cite{Emil2020scalable}.
The estimation method is detailed in \cite[(3.27)]{Emil2017mMnetworks}, where the $\vect{R}_{kl}$ (to be estimated) is formed as a convex combination of a diagonal matrix and a sample matrix for optimizing the weighting parameter to achieve a robust estimation.
In addition, $\beta_{kl}^\mathrm{NLOS}=\beta_{kl}/(\kappa_{kl}+1)$ is the large-scale fading of the NLOS path where $\beta_{kl}$ and $\kappa_{kl}$ denote the received power of the faded signal and Rician factor between $k^\mathrm{th}$ UE and $l^\mathrm{th}$ AP, respectively.
Assuming a half-wavelength uniform linear array \cite{Q.Zhang2014a}, we have
$
\vect{\bar{h}}_{kl}=\sqrt{\beta_{kl}^\mathrm{LOS}}[1,\exp({-\mathrm{j}\pi\sin(\theta_{kl})}),\cdots,\exp({-\mathrm{j}(N-1)\pi\sin(\theta_{kl})})]^{\mathrm{T}}
$
where $\theta_{kl}$ and $\beta_{kl}^\mathrm{LOS}=\kappa_{kl}\beta_{kl}/(\kappa_{kl}+1)$ are the angle of arrival (AoA) and the large-scale fading of the LOS path between $k^\mathrm{th}$ UE and $l^\mathrm{th}$ AP, respectively. Similar to \cite{Ozdogan2019Massive}, the AoA of channel LOS component is also assumed to be known in our system and the detailed estimation method using IQ-value is discussed in \cite{Iwamoto2017}.


\subsection{Quantization Model for Finite Precision DACs/ADCs}
As shown in Fig. \ref{fig_system_model}(b), each UE and AP employ with finite precision DACs and ADCs, respectively. For the input signal vector $\vect{x}$, the quantized signal can be expressed as \cite{Xu2019ADC,Xu2019DAC}
\begin{align}
&f_\mathrm{da}(\vect{x})=\sqrt{1-\rho_\mathrm{da}}\vect{x}+\vect{n}_\mathrm{da},\label{f_dac}
\\&f_\mathrm{ad}(\vect{x})=(1-\rho_\mathrm{ad})\vect{x}+\vect{n}_\mathrm{ad},\label{f_adc}
\end{align}
with $\vect{n}_\mathrm{da}\sim\mathcal{CN}(\vect{0},\vect{C}_\mathrm{da})$ and $\vect{n}_\mathrm{ad}\sim\mathcal{CN}(\vect{0},\vect{C}_\mathrm{ad})$ denoting the quantization noise caused by finite precision DACs and ADCs, respectively, where
\begin{align}
&\vect{C}_\mathrm{da}=\rho_\mathrm{da}\diag(\mathbb{E}[\vect{x}\vect{x}^\mathrm{H}]),\label{dac}
\\&\vect{C}_\mathrm{ad}=\rho_\mathrm{ad}(1-\rho_\mathrm{ad})\diag(\mathbb{E}[\vect{x}\vect{x}^\mathrm{H}]).\label{adc}
\end{align}
Notice that, although $\vect{n}_\mathrm{da}$ and $\vect{n}_\mathrm{ad}$ depend only on the statistics of $\vect{x}$, they are independent to the random vector $\vect{x}$ \cite{Xu2019ADC,Xu2019DAC}.
For small integer number of quantization bits (Q-bits) $b_\mathrm{da/ad}$,  the distortion factor $\rho_\mathrm{da/ad}$ takes the typical values listed in Table \ref{table_coefficients}. For $b_\mathrm{da/ad}>5$, however, its value can be approximated as $\rho_\mathrm{da/ad}\approx\sqrt{3}\pi2^{-2b_\mathrm{da/ad}-1}$ \cite{J.Zhang2015Mix}.
\begin{table}[t]
\caption{\textsc{Values of $\rho_\mathrm{da/ad}$ for Small $b_\mathrm{da/ad}$} }
\vskip 1mm
\label{table_coefficients}
\centering
\begin{tabular}{c||ccccc}
\hline
\bfseries \quad $b_\mathrm{da/ad}$&1&2&3&4&5\\
\hline
\bfseries \quad $\rho_\mathrm{da/ad}$&0.3634&0.1175&0.03454&0.009497&0.002499\\
\hline
\end{tabular}\vspace{-0.5cm}
\end{table}

\subsection{MMSE Channel Estimation}
For orthogonal pilots, each UE's pilot can be taken from the column of a $\tau\times\tau$ discrete Fourier transform (DFT) matrix \cite[(3.8)]{Emil2017mMnetworks}
where each column satisfies $\|\bm{\phi}_{t_1}\|^2=\tau$ with $[\bm{\phi}_{t_1}]_{t_2}=\exp[-\mathrm{j}2\pi({t_1}-1)({t_2}-1)/\tau]$.
Since the high connectivity scenario is considered for which $K>\tau$ and denoting the pilot index assigned to $k^\mathrm{th}$ UE as $t_k^\mathrm{p}\in\{1,2,\cdots,\tau\}$, $\mathcal{P}_k\subset\{1,2,\cdots,K\}$ is the subset of UEs assigned to pilot $t_k^\mathrm{p}$.
According to the transceiver model illustrated in Fig. \ref{fig_system_model}(b), after the transmit pilots of each UE pass through the finite precision DACs and ADCs, the quantized received pilot matrix at $l^\mathrm{th}$ AP, $\vect{Z}_l\in\mathbb{C}^{N\times\tau}$, can be expressed as
\begin{align}\notag
\vect{Z}_l&=(1-\rho_\mathrm{ad})\left[\sum_{i=1}^{K}\vect{h}_{il}\big(\sqrt{\ddot{p}_i}\bm{\phi}_{t_i^\mathrm{p}}^\mathrm{T}+\vect{n}_{\mathrm{da},i}^\mathrm{T}\big)
+\vect{N}_{\mathrm{an},l}\right]+\vect{N}_{\mathrm{ad},l}
\\&=(1-\rho_\mathrm{ad})\sum_{i=1}^{K}\sqrt{\ddot{p}_i}\vect{h}_{il}\bm{\phi}_{t_i^\mathrm{p}}^\mathrm{T}+\vect{N}_{l},
\end{align}
where $\ddot{p}_i=(1-\rho_\mathrm{da})p_i$ and $p_i$ is the transmit power of $i^\mathrm{th}$ UE, $\vect{n}_{\mathrm{da},i}\sim\mathcal{CN}(\vect{0},\rho_\mathrm{da}p_i\vect{I}_\tau)$ is transmit noise of $i^\mathrm{th}$ UE caused by finite resolution DACs,
$\vect{N}_{\mathrm{an},l}$ is the received noise at $l^\mathrm{th}$ AP with i.i.d. element $[\vect{N}_{\mathrm{an},l}]_{ij}\sim\mathcal{CN}(0,\sigma^2)$, and $\vect{N}_{l}=[\vect{n}_{l1},\vect{n}_{l2},\cdots,\vect{n}_{l\tau}]$ is given as
\begin{equation}\label{Nl}
\vect{N}_{l}=(1-\rho_\mathrm{ad})\sum_{i=1}^{K}\vect{h}_{il}\vect{n}_{\mathrm{da},i}^\mathrm{T}
+(1-\rho_\mathrm{ad})\vect{N}_{\mathrm{an},l}+\vect{N}_{\mathrm{ad},l}.
\end{equation}
To obtain $\vect{\hat{h}}_{kl}$, $l^\mathrm{th}$ AP fist correlates $\vect{Z}_l$ with $\bm{\phi}_{t_k^\mathrm{p}}^\ast/\sqrt{\tau}$ which gives
\begin{align}\label{zp}
\vect{z}_{t_k^\mathrm{p},l}&=\vect{Z}_l\bm{\phi}_{t_k^\mathrm{p}}^\ast/\sqrt{\tau}=(1-\rho_\mathrm{ad})\sum_{i\in\mathcal{P}_k}\sqrt{\ddot{p}_i\tau}\vect{h}_{il}+\vect{n}_{t_k^\mathrm{p},l},
\end{align}
where $\vect{n}_{t_k^\mathrm{p},l}=\vect{N}_{l}\bm{\phi}_{t_k^\mathrm{p}}^\ast/\sqrt{\tau}$. For $t=1,2,\cdots,\tau$, by using \eqref{dac} and \eqref{adc}, $\vect{C}_{\mathrm{n},lt}=\mathbb{E}[\vect{n}_{lt}\vect{n}_{lt}^\mathrm{H}]$ is given as
\begin{align}
\notag\vect{C}_{\mathrm{n},lt}&=(1-\rho_\mathrm{ad})^2\rho_\mathrm{da}\mathbb{E}[\vect{H}_l\vect{P}\vect{H}_l^\mathrm{H}]+(1-\rho_\mathrm{ad})\sigma^2\vect{I}_N
\\&\quad+\rho_\mathrm{ad}(1-\rho_\mathrm{ad})\diag(\mathbb{E}[\vect{H}_l\vect{P}\vect{H}_l^\mathrm{H}]),
\end{align}
where $\vect{H}_l=[\vect{h}_{1l},\vect{h}_{2l},\cdots,\vect{h}_{Kl}]$ and $\vect{P}\triangleq[p_i]_{K}^\Lambda$. Since
$
\mathbb{E}[\vect{H}_l\vect{P}\vect{H}_l^\mathrm{H}]=\vect{\bar{H}}_l\vect{P}\vect{\bar{H}}_l^\mathrm{H}+\sum_{i=1}^{K}{p_i}\vect{R}_{il} $
with $\vect{\bar{H}}_l=[\vect{\bar{h}}_{1l},\vect{\bar{h}}_{2l},\cdots,\vect{\bar{h}}_{Kl}]$, $\vect{C}_{\mathrm{n},lt}$ can be rewritten as
\begin{align}
\notag\vect{C}_{\mathrm{n},lt}&=\notag\vect{C}_{\mathrm{n},l}=\frac{(1-\rho_\mathrm{ad})^2\rho_\mathrm{da}}{1-\rho_\mathrm{da}}\bigg(\vect{\bar{H}}_l\vect{\ddot{P}}\vect{\bar{H}}_l^\mathrm{H}+\sum_{i=1}^{K}{\ddot{p}_i}\vect{R}_{il}\bigg)
\\\notag&+\frac{\rho_\mathrm{ad}(1-\rho_\mathrm{ad})}{1-\rho_\mathrm{da}}\diag\bigg(\vect{\bar{H}}_l\vect{\ddot{P}}\vect{\bar{H}}_l^\mathrm{H}+\sum_{i=1}^{K}{\ddot{p}_i}\vect{R}_{il}\bigg)
\\&+(1-\rho_\mathrm{ad})\sigma^2\vect{I}_N
\end{align}
with $\vect{\ddot{P}}\triangleq[\ddot{p}_i]_{K}^\Lambda$, which is independent of $t$. Observing \eqref{Nl}, it is clear that
$\vect{n}_{l,t_1}$ and $\vect{n}_{l,t_2}$ are independent for $t_1\neq t_2$, thus one can obtain that $\vect{n}_{t_k^\mathrm{p},l}=\vect{N}_{l}\bm{\phi}_{t_k^\mathrm{p}}^\ast/\sqrt{\tau}\sim\mathcal{CN}(\vect{0},\vect{C}_{\mathrm{n},l})$. In addition, since $\bm{\phi}_{t_k^\mathrm{p}}^\mathrm{T}\bm{\phi}_{t_i^\mathrm{p}}=0$ for $i\notin\mathcal{P}_k$, one can obtain that $\mathbb{E}[\vect{n}_{t_k^\mathrm{p},l}\vect{n}_{t_i^\mathrm{p},l}^\mathrm{H}]=\vect{0}$ for $i\notin\mathcal{P}_k$.
Applying \cite[Corollary B.$18$]{Emil2017mMnetworks} to \eqref{zp}, the MMSE estimate of $\vect{h}_{kl}$ is \cite{J.Zhang2020NOMA}
\begin{equation}\label{hhat}
\vect{\hat{h}}_{kl}=\vect{\bar{h}}_{kl}+(1-\rho_\mathrm{ad})\sqrt{\ddot{p}_k\tau}\vect{R}_{kl}\vect{\Psi}_{t_k^\mathrm{p},l}^{-1}\vect{z}_{t_k^\mathrm{p},l}^\mathrm{w},
\end{equation}
where $\vect{z}_{t_k^\mathrm{p},l}^\mathrm{w}=(1-\rho_\mathrm{ad})\sum_{i\in\mathcal{P}_k}\sqrt{\ddot{p}_i\tau}\vect{h}_{\mathrm{w},il}+\vect{n}_{t_k^\mathrm{p},l}$ and
\begin{equation}\label{phi_tp}
\vect{\Psi}_{t_k^\mathrm{p},l}=\mathbb{E}\big[\vect{z}_{t_k^\mathrm{p},l}^\mathrm{w}\big(\vect{z}_{t_k^\mathrm{p},l}^\mathrm{w}\big)^\mathrm{H}\big]=(1-\rho_\mathrm{ad})^2\sum_{i\in\mathcal{P}_k}\ddot{p}_i\tau\vect{R}_{il}+\vect{C}_{\mathrm{n},l}. \end{equation}
Thus, $\vect{\hat{h}}_{kl}$ and its estimation error, $\vect{\tilde{h}}_{kl}=\vect{h}_{kl}-\vect{\hat{h}}_{kl}$, are mutually independent distributed as
\begin{align}
&\vect{\hat{h}}_{kl}\sim\mathcal{CN}\big(\vect{\bar{h}}_{kl},\vect{C}_{\vect{\hat{h}}_{kl}}\big),\label{h_hat_D}
\\&\vect{\tilde{h}}_{kl}\sim\mathcal{CN}\big(\vect{0},\vect{R}_{kl}-\vect{C}_{\vect{\hat{h}}_{kl}}\big),\label{h_tilde_D}
\end{align}
where $\vect{C}_{\vect{\hat{h}}_{kl}}=(1-\rho_\mathrm{ad})^2\ddot{p}_k\tau
\vect{R}_{kl}\vect{\Psi}_{t_k^\mathrm{p},l}^{-1}\vect{R}_{kl}$.
\subsection{Data Signal Transmission for the Distributed Scheme}
Similarly to the pilot training phase, the quantized received data signal vector at $l^\mathrm{th}$ AP, $\vect{y}_l\in\mathbb{C}^{N}$, can be expressed as
\begin{align}\label{y_l_D}
\notag\vect{y}_l&=(1-\rho_\mathrm{ad})\sum_{i=1}^{K}\vect{h}_{il}(\ddot{s}_{i}+n_{\mathrm{da},i})
+(1-\rho_\mathrm{ad})\vect{n}_{\mathrm{an},l}+\vect{n}_{\mathrm{ad},l}
\\&=(1-\rho_\mathrm{ad})\sum_{i=1}^{K}\vect{h}_{il}\ddot{s}_{i}+\vect{n}_{l},
\end{align}
where $\ddot{s}_i=\sqrt{1-\rho_\mathrm{da}}s_i$ with $s_i\sim\mathcal{CN}(0,p_i)$ be the transmit signal of the $i^\mathrm{th}$ UE, and
\begin{equation}\label{n_l-a}
\vect{n}_{l}=(1-\rho_\mathrm{ad})\sum_{i=1}^{K}\vect{h}_{il}n_{\mathrm{da},i}
+(1-\rho_\mathrm{ad})\vect{n}_{\mathrm{an},l}+\vect{n}_{\mathrm{ad},l}
\end{equation}
with $\vect{n}_{l}\sim\mathcal{CN}(\vect{0},\vect{C}_{\mathrm{n},l})$.
Letting $\mathcal{M}_k\subset\{1,2,\cdots,L\}$ be the APs cluster serving $k^\mathrm{th}$ UE \cite{Emil2020scalable}, the $l^\mathrm{th}$ AP ($l\in\mathcal{M}_k$) can locally detect the data of $k^\mathrm{th}$ UE from $\vect{y}_l$ by using the combining vector $\vect{v}_{kl}\in\mathbb{C}^N$ depending on the local estimated channel $\vect{\hat{H}}_l$, which yields
\begin{equation}
\hat {s}_{kl}^\mathrm{d}=\vect{v}_{kl}^\mathrm{H}\vect{y}_l=(1-\rho_\mathrm{ad})\vect{v}_{kl}^\mathrm{H}\vect{h}_{kl}\ddot{s}_{k}
+(1-\rho_\mathrm{ad})\vect{v}_{kl}^\mathrm{H}\sum_{i\neq k}^{K}\vect{h}_{il}\ddot{s}_{i}
+\vect{v}_{kl}^\mathrm{H}\vect{n}_{l}.
\end{equation}
According to \cite[($16$)]{Emil2020making}, the local MMSE (L-MMSE) combing vector which minimizes $\mathbb{E}\big[|s_{kl}-\hat{s}_{kl}^\mathrm{d}|^2|\{\vect{\hat{h}}_{il}\}\big]$ is given as
\begin{align}\label{L-MMSE_detector}
\notag&\vect{v}_{kl}^\mathrm{L-MMSE}=
\\&\bigg[(1-\rho_\mathrm{ad})^2\sum_{i=1}^{K}\ddot{p}_i\big(\vect{\hat{h}}_{il}\vect{\hat{h}}_{il}^\mathrm{H}+\vect{R}_{il}
-\vect{C}_{\vect{\hat{h}}_{il}}\big)
+\vect{C}_{\mathrm{n},l}\bigg]^{-1}\vect{\hat{h}}_{kl},
\end{align}
which is an unscalable detector because the total number of complex multiplications required to calculate $\sum_{i=1}^{K}\ddot{p}_i(\vect{\hat{h}}_{il}\vect{\hat{h}}_{il}^\mathrm{H}+\vect{R}_{il}
-\vect{C}_{\vect{\hat{h}}_{il}})$ in \eqref{L-MMSE_detector} grows linearly with $K$.
As an improvement, the following scalable LP-MMSE vector, $\vect{v}_{k}^\mathrm{LP-MMSE}$, is proposed:
\begin{align}\label{LP-MMSE_detector}
\notag&\vect{v}_{kl}^\mathrm{LP-MMSE}=
\Bigg[(1-\rho_\mathrm{ad})^2\sum_{i\in\mathcal{N}_l^\mathrm{P}}\ddot{p}_i\big(\vect{\hat{h}}_{il}\vect{\hat{h}}_{il}^\mathrm{H}+\vect{R}_{il}
-\vect{C}_{\vect{\hat{h}}_{il}}\big)
\\\notag&+(1-\rho_\mathrm{ad})^2\sum_{i\in\mathcal{N}_l^\mathrm{S}}\ddot{p}_i\big(\vect{\bar{h}}_{il}\vect{\bar{h}}_{il}^\mathrm{H}+\vect{R}_{il}\big)
+(1-\rho_\mathrm{ad})\sigma^2\vect{I}_N
\\\notag&+\frac{(1-\rho_\mathrm{ad})^2\rho_\mathrm{da}}{1-\rho_\mathrm{da}}\sum_{i\in\mathcal{N}_l}{\ddot{p}_i}\big(\vect{\bar{h}}_{il}\vect{\bar{h}}_{il}^\mathrm{H}+\vect{R}_{il}\big)
\\&+\frac{\rho_\mathrm{ad}(1-\rho_\mathrm{ad})}{1-\rho_\mathrm{da}}\diag\bigg(\sum_{i\in\mathcal{N}_l}{\ddot{p}_i}\big(\vect{\bar{h}}_{il}\vect{\bar{h}}_{il}^\mathrm{H}+\vect{R}_{il}\big)\bigg)
\Bigg]^{-1}\vect{\hat{h}}_{kl},
\end{align}
where $\mathcal{N}_l=\{i|\vect{D}_{il}=\vect{I}_N,i\in\{1,2,\cdots,K\}\}$ denotes the UE set served by $l^\mathrm{th}$ AP with $\vect{D}_{il}=\vect{I}_N$ if $l\in\mathcal{M}_i$ and $\vect{0}_N$ otherwise; $\mathcal{N}_l^\mathrm{P}\subset\mathcal{N}_l$ denotes the UE set where $l^\mathrm{th}$ AP is assigned as their primary AP (P-AP); and $\mathcal{N}_l^\mathrm{S}\subset\mathcal{N}_l$ denotes the UE set where $l^\mathrm{th}$ AP is assigned as their sub-AP (S-AP). It is noted that $\mathcal{N}_l^\mathrm{P}\cup\mathcal{N}_l^\mathrm{S}=\mathcal{N}_l$ and $\mathcal{N}_l^\mathrm{P}\cap\mathcal{N}_l^\mathrm{S}=\emptyset$.
The exact procedure of how $\mathcal{N}_l$ in \eqref{LP-MMSE_detector} is generated will be presented in Section \ref{section:algorithm} where the joint AP cluster formation, pilot assignment and UL power control algorithm (termed as Algorithm \ref{alg1}) can be found.
Compared with the LP-MMSE vector in \cite{Emil2020scalable} and \cite{Chen2020Structured} which have been proposed for the Rayleigh fading, there are two main novelties offered by our proposed LP-MMSE vector: \textit{i}) It is more general as it includes of the Rician factor and the quantization noise caused by the finite resolution of ADCs and DACs; \textit{ii}) It requires significantly less CC\footnote{It is noted that a comparative CC analysis for the different scalable MMSE detectors will be presented later on in Section \ref{cc}.}, while it turns out that the corresponding SE performance is similar to that of the original LP-MMSE vector in \cite{Emil2020scalable}, which will be presented later on in Fig. \ref{UL-SE_CDF}(a).

The local data estimates $\{\hat {s}_{kl}^\mathrm{d}|l\in\mathcal{M}_k\}$ are sent to CPUs to finally detect $s_k^\mathrm{d}$ as follows
\begin{align}
\notag\hat{s}_k^\mathrm{d}&=\sum_{l\in\mathcal{M}_k}a_{kl}^\ast\hat {s}_{kl}^\mathrm{d}
=(1-\rho_\mathrm{ad})\sum_{l\in\mathcal{M}_k}a_{kl}^\ast\vect{v}_{kl}^\mathrm{H}\vect{h}_{kl}\ddot{s}_{k}
\\\notag&\quad+(1-\rho_\mathrm{ad})\sum_{l\in\mathcal{M}_k}a_{kl}^\ast\vect{v}_{kl}^\mathrm{H}\sum_{i\neq k}^{K}\vect{h}_{il}\ddot{s}_{i}
+\sum_{l\in\mathcal{M}_k}a_{kl}^\ast\vect{v}_{kl}^\mathrm{H}\vect{n}_{l}
\\&=(1-\rho_\mathrm{ad})\vect{a}_k^\mathrm{H}\vect{g}_{kk}\ddot{s}_k+(1-\rho_\mathrm{ad})\sum_{i\neq k}^{K}\vect{a}_k^\mathrm{H}\vect{g}_{ki}\ddot{s}_i+n_k^\prime
\end{align}
where
$\vect{a}_k=[a_{kl}|l\in\mathcal{M}_k]^\mathrm{T}$ denotes the large-scale fading decoding (LSFD) vector, $\vect{g}_{ki}=[\vect{v}_{kl}^\mathrm{H}\vect{h}_{il}|l\in\mathcal{M}_k]^\mathrm{T}$,
and $n_k^\prime=\sum_{l\in\mathcal{M}_k}a_{kl}^\ast\vect{v}_{kl}^\mathrm{H}\vect{n}_{l}$.
Since the CPUs do not have knowledge of the channel estimates, they consider the average channel $(1-\rho_\mathrm{ad})\vect{a}_k^\mathrm{H}\mathbb{E}[\vect{g}_{kk}]$ as the actual channel. Hence the signal model can be reexpressed as \cite{Emil2020making},\cite[(4.13)]{Emil2017mMnetworks}
\begin{align}\label{eq:UL-d}
\notag\hat{s}_k^\mathrm{d}&=(1-\rho_\mathrm{ad})\vect{a}_k^\mathrm{H}\mathbb{E}[\vect{g}_{kk}]\ddot{s}_k
+(1-\rho_\mathrm{ad})\vect{a}_k^\mathrm{H}(\vect{g}_{kk}-\mathbb{E}[\vect{g}_{kk}])\ddot{s}_k
\\\notag&\quad+(1-\rho_\mathrm{ad})\sum_{i\neq k}^{K}\vect{a}_k^\mathrm{H}\vect{g}_{ki}\ddot{s}_i
\\&=(1-\rho_\mathrm{ad})\vect{a}_k^\mathrm{H}\mathbb{E}[\vect{g}_{kk}]\ddot{s}_k+n_k^{\prime\prime},
\end{align}
where $n_k^{\prime\prime}=(1-\rho_\mathrm{ad})\big[\vect{a}_k^\mathrm{H}(\vect{g}_{kk}-\mathbb{E}[\vect{g}_{kk}])\ddot{s}_k
+\sum_{i\neq k}^{K}\vect{a}_k^\mathrm{H}\vect{g}_{ki}\ddot{s}_i\big]+n_k^\prime$ has zero mean and is uncorrelated with the signal term $(1-\rho_\mathrm{ad})\vect{a}_k^\mathrm{H}\mathbb{E}[\vect{g}_{kk}]\ddot{s}_k$. From \eqref{eq:UL-d}, the UL-SE of $k^\mathrm{th}$ UE can be expressed as \eqref{UL-SE}, where $\vect{f}_k=[\vect{v}_{kl}^\mathrm{H}\vect{n}_{l}|l\in\mathcal{M}_k]^\mathrm{T}$.
\begin{figure*}[!t]
\begin{equation}\label{UL-SE}
R_{k}^\mathrm{d}=\Big(1-\frac{\tau}{\tau_c}\Big)\log_2\Bigg(1+\frac{(1-\rho_\mathrm{ad})^2\ddot{p}_k|\vect{a}_k^\mathrm{H}\mathbb{E}[\vect{g}_{kk}]|^2}
{(1-\rho_\mathrm{ad})^2\sum_{i=1}^{K}\ddot{p}_i\mathbb{E}[|\vect{a}_k^\mathrm{H}\vect{g}_{ki}|^2]-(1-\rho_\mathrm{ad})^2\ddot{p}_k|\vect{a}_k^\mathrm{H}\mathbb{E}[\vect{g}_{kk}]|^2+\vect{a}_k^\mathrm{H}\mathbb{E}[\vect{f}_k\vect{f}_k^\mathrm{H}]\vect{a}_k}\Bigg)
\end{equation}
\hrulefill
\end{figure*}
Note that the signal-to-interference-plus-noise ratio (SINR) term in \eqref{UL-SE} can be expressed as a generalized Rayleigh quotient form as
\begin{align}\label{SNR-d}
\left\{
\begin{array}{lr}
\mathrm{SINR}_{k}^\mathrm{d}=(1-\rho_\mathrm{ad})^2\ddot{p}_k{|\vect{a}_k^\mathrm{H}\mathbb{E}[\vect{g}_{kk}]|^2}/({\vect{a}_k^\mathrm{H}\vect{B}_k^\mathrm{d}\vect{a}_k}),
\\\vect{B}_k^\mathrm{d}=(1-\rho_\mathrm{ad})^2\sum_{i=1}^{K}\ddot{p}_i\mathbb{E}[\vect{g}_{ki}\vect{g}_{ki}^\mathrm{H}]
+\mathbb{E}[\vect{f}_k\vect{f}_k^\mathrm{H}]
\\\qquad\;\;-(1-\rho_\mathrm{ad})^2\ddot{p}_k\mathbb{E}[\vect{g}_{kk}]\mathbb{E}[\vect{g}_{kk}^\mathrm{H}].
\end{array}
\right.
\end{align}
Thus the optimal LSFD vector, $\vect{a}_k^\mathrm{opt}$, is given as
\begin{align}\label{LSFD}
\notag&\vect{a}_k^\mathrm{opt}=(\vect{B}_k^\mathrm{d})^{-1}\mathbb{E}[\vect{g}_{kk}]=\bigg[(1-\rho_\mathrm{ad})^2\sum_{i=1}^{K}\ddot{p}_i\mathbb{E}[\vect{g}_{ki}\vect{g}_{ki}^\mathrm{H}]
\\&+\mathbb{E}[\vect{f}_k\vect{f}_k^\mathrm{H}]
-(1-\rho_\mathrm{ad})^2\ddot{p}_k\mathbb{E}[\vect{g}_{kk}]\mathbb{E}[\vect{g}_{kk}^\mathrm{H}]\bigg]^{-1}\mathbb{E}[\vect{g}_{kk}],
\end{align}
which yields $\mathrm{SINR}_{k,\max}^\mathrm{d}=(1-\rho_\mathrm{ad})^2\ddot{p}_k\mathbb{E}[\vect{g}_{kk}^\mathrm{H}](\vect{B}_k^\mathrm{d})^{-1}\mathbb{E}[\vect{g}_{kk}]$. Then the maximum value of $R_{k}^\mathrm{d}$ in \eqref{UL-SE} is given in \eqref{UL-SE-max}.
\begin{figure*}[!t]
\begin{align}\label{UL-SE-max} R_{k,\max}^\mathrm{d}=\Big(1-\frac{\tau}{\tau_c}\Big)\log_2\Bigg(1+(1-\rho_\mathrm{ad})^2\ddot{p}_k\mathbb{E}[\vect{g}_{kk}^\mathrm{H}]&
\bigg[(1-\rho_\mathrm{ad})^2\sum_{i=1}^{K}\ddot{p}_i\mathbb{E}[\vect{g}_{ki}\vect{g}_{ki}^\mathrm{H}]
+\mathbb{E}[\vect{f}_k\vect{f}_k^\mathrm{H}]
-(1-\rho_\mathrm{ad})^2\ddot{p}_k\mathbb{E}[\vect{g}_{kk}]\mathbb{E}[\vect{g}_{kk}^\mathrm{H}]\bigg]^{-1}\mathbb{E}[\vect{g}_{kk}]\Bigg)
\end{align}
\hrulefill
\end{figure*}
\subsection{Data Signal Transmission for the Centralized Scheme}
Different from distributed scheme, for the centralized scheme case $L$ APs forward their received pilot signals $\{\vect{z}_{t^\mathrm{p},l}|t^\mathrm{p}=1,2,\cdots,\tau;l=1,2,\cdots,L\}$ and data signals $\{\vect{y}_{l}|l=1,2,\cdots,L\}$ to CPUs where the CE and data signal detection are carried out in a centralized manner.
By letting $\vect{\hat{h}}_k=[\vect{\hat{h}}_{k1}^\mathrm{T},\vect{\hat{h}}_{k2}^\mathrm{T},\cdots,\vect{\hat{h}}_{kL}^\mathrm{T}]^\mathrm{T}$, and $\vect{\bar{h}}_k=[\vect{\bar{h}}_{k1}^\mathrm{T},\vect{\bar{h}}_{k2}^\mathrm{T},\cdots,\vect{\bar{h}}_{kL}^\mathrm{T}]^\mathrm{T}$, similarly to \eqref{hhat}, the $k^\mathrm{th}$ UE's MMSE CE computed at the CPU can be expressed as follows
\begin{equation}\label{central_h_hat}
\vect{\hat{h}}_k=\vect{\bar{h}}_{k}+(1-\rho_\mathrm{ad})\sqrt{\ddot{p}_k\tau}\big[\vect{R}_{kl}\vect{\Psi}_{t_k^\mathrm{p},l}^{-1}\big]_{L}^\mathrm{B}\vect{z}_{t_k^\mathrm{p}}^\mathrm{w},
\end{equation}
where $\vect{z}_{t_k^\mathrm{p}}^\mathrm{w}=\big[\big(\vect{z}_{t_k^\mathrm{p},1}^\mathrm{w}\big)^\mathrm{T},\big(\vect{z}_{t_k^\mathrm{p},2}^\mathrm{w}\big)^\mathrm{T},\cdots,\big(\vect{z}_{t_k^\mathrm{p},L}^\mathrm{w}\big)^\mathrm{T}\big]^\mathrm{T}\sim\mathcal{CN}(\vect{0},[\vect{\Psi}_{t_k^\mathrm{p},l}]_{L}^\mathrm{B})$. Defining $\vect{h}_k=[\vect{h}_{k1}^\mathrm{T},\vect{h}_{k2}^\mathrm{T},\cdots,\vect{h}_{kL}^\mathrm{T}]^\mathrm{T}$,
similarly to \eqref{h_hat_D} and \eqref{h_tilde_D}, $\vect{\hat{h}}_k$ and its CE error, $\vect{\tilde{h}}_{k}=\vect{h}_{k}-\vect{\hat{h}}_{k}$, follow the distribution as
\begin{align}
&\vect{\hat{h}}_k
\sim\mathcal{CN}\big(\vect{\bar{h}}_{k},\vect{C}_{\vect{\hat{h}}_k}\big),\label{h-hat-sim}
\\&\vect{\tilde{h}}_{k}\sim\mathcal{CN}\big(\vect{0},[\vect{R}_{kl}]_{L}^\mathrm{B}
-\vect{C}_{\vect{\hat{h}}_k}\big),\label{h-tilde-sim}
\end{align}
respectively, where $\vect{C}_{\vect{\hat{h}}_k}=(1-\rho_\mathrm{ad})^2\ddot{p}_k\tau
\big[\vect{R}_{kl}\vect{\Psi}_{t_k^\mathrm{p},l}^{-1}\vect{R}_{kl}\big]_{L}^\mathrm{B}$.
The block-diagonal form of the covariance matrix above show that $k^\mathrm{th}$ UE's CE vectors at different AP are mutually independent.

Similar to \eqref{y_l_D}, during the UL data transmission phase, the received signal vector at CPUs can be expressed as
\begin{equation}
\vect{y}=(1-\rho_\mathrm{ad})\sum_{i=1}^{K}\vect{h}_i\ddot{s}_i+\vect{n},
\end{equation}
where
$\vect{n}=[\vect{n}_{1}^\mathrm{T},\vect{n}_{2}^\mathrm{T},\cdots,\vect{n}_{L}^\mathrm{T}]^\mathrm{T}
\sim\mathcal{CN}(\vect{0},[\vect{C}_{\mathrm{n},l}]_{L}^\mathrm{B})$.
Then the CPUs detect from $\vect{y}$ the data of $k^\mathrm{th}$ UE by using the detection vector $\vect{D}_k\vect{v}_{k}=\big[\vect{D}_{kl}\big]_{L}^\mathrm{B}[\vect{v}_{k1}^\mathrm{T},\vect{v}_{k2}^\mathrm{T},\cdots,\vect{v}_{kL}^\mathrm{T}]^\mathrm{T}$,
which gives
\begin{align}\label{eq:UL-c}
\notag\hat {s}_{k}^\mathrm{c}&=\vect{v}_{k}^\mathrm{H}\vect{D}_k\vect{y}=(1-\rho_\mathrm{ad})\vect{v}_{k}^\mathrm{H}\vect{D}_k\vect{\hat{h}}_k\ddot{s}_k
+\vect{v}_{k}^\mathrm{H}\vect{D}_k\vect{n}
\\&+(1-\rho_\mathrm{ad})\Bigg(\sum_{i\neq k}^{K}\vect{v}_{k}^\mathrm{H}\vect{D}_k\vect{\hat{h}}_i\ddot{s}_i
+\sum_{i=1}^{K}\vect{v}_{k}^\mathrm{H}\vect{D}_k\vect{\tilde{h}}_i\ddot{s}_i\Bigg).
\end{align}
Using \eqref{eq:UL-c}, the UL-SE of $k^\mathrm{th}$ UE can be expressed as \eqref{UL-SE-C}.
\begin{figure*}[!t]
\begin{align}\label{UL-SE-C}
R_{k}^\mathrm{c}=\Big(1-\frac{\tau}{\tau_c}\Big)
\mathbb{E}\Bigg[\log_2\Bigg(1+
\frac{(1-\rho_\mathrm{ad})^2\ddot{p}_k|\vect{v}_{k}^\mathrm{H}\vect{D}_k\vect{\hat{h}}_k|^2}
{(1-\rho_\mathrm{ad})^2\sum_{i\neq k}^{K}\ddot{p}_i|\vect{v}_{k}^\mathrm{H}\vect{D}_k\vect{\hat{h}}_i|^2
+\vect{v}_{k}^\mathrm{H}\vect{D}_k\big[(1-\rho_\mathrm{ad})^2\sum_{i=1}^{K}\ddot{p}_i\big([\vect{R}_{il}]_{L}^\mathrm{B}
-\vect{C}_{\vect{\hat{h}}_i}\big)+[\vect{C}_{\mathrm{n},l}]_{L}^\mathrm{B}\big]\vect{D}_k\vect{v}_{k}}\Bigg)\Bigg]
\end{align}
\hrulefill
\end{figure*}
Similarly to \eqref{UL-SE}, the instantaneous SINR of \eqref{UL-SE-C} can be expressed as a generalized Rayleigh quotient which is maximized by $\vect{v}_{k}^\mathrm{opt}=(\vect{B}_k^\mathrm{c})^\dagger\vect{D}_k\vect{\hat{h}}_k$, which leads to
\begin{align}
\left\{
\begin{array}{lr}
\mathrm{SINR}_{k,\max}^\mathrm{c}=(1-\rho_\mathrm{ad})^2\ddot{p}_k\vect{\hat{h}}_k^\mathrm{H}\vect{D}_k(\vect{B}_k^\mathrm{c})^\dagger\vect{D}_k\vect{\hat{h}}_k,
\\\vect{B}_k^\mathrm{c}=\vect{D}_k\bigg[\sum_{i\neq k}^{K}(1-\rho_\mathrm{ad})^2\ddot{p}_i\vect{\hat{h}}_i\vect{\hat{h}}_i^\mathrm{H}
+[\vect{C}_{\mathrm{n},l}]_{L}^\mathrm{B}
\\+(1-\rho_\mathrm{ad})^2\sum_{i=1}^{K}\ddot{p}_i\big([\vect{R}_{il}]_{L}^\mathrm{B}
-\vect{C}_{\vect{\hat{h}}_i}\big)\bigg]\vect{D}_k
\end{array}
\right.
\end{align}
According to \cite[(C.$51$)]{Emil2017mMnetworks}, $\mathrm{SINR}_{k,\max}^\mathrm{c}$ can also be achieved by using the MMSE combining vector \cite[Corollary $1$]{Emil2020scalable}
\begin{equation}\label{MMSE_detector}
\vect{v}_{k}^\mathrm{MMSE}=(\vect{B}_k^\mathrm{c}+(1-\rho_\mathrm{ad})^2\ddot{p}_k\vect{D}_k\vect{\hat{h}}_k\vect{\hat{h}}_k^\mathrm{H}\vect{D}_k)^\dagger\vect{D}_k\vect{\hat{h}}_k,
\end{equation}
which is an unscalable detector because the total number of complex multiplications required to calculate $\sum_{i=1}^{K}\ddot{p}_i\vect{D}_k\vect{\hat{h}}_i\vect{\hat{h}}_i^\mathrm{H}\vect{D}_k$ and $\sum_{i=1}^{K}\ddot{p}_i\vect{D}_k\big([\vect{R}_{il}]_{L}^\mathrm{B}
-\vect{C}_{\vect{\hat{h}}_i}\big)\vect{D}_k$ in \eqref{MMSE_detector} grows with $K$.
Recalling the fact that the interference affecting $k^\mathrm{th}$ UE is mainly caused by a small set of nearby UEs, only the UE whose AP cluster partially overlaps with the AP cluster of $k^\mathrm{th}$ UE does should be included into the summation terms of $\sum_{i=1}^{K}\ddot{p}_i\vect{D}_k\vect{\hat{h}}_i\vect{\hat{h}}_i^\mathrm{H}\vect{D}_k$ and $\sum_{i=1}^{K}\ddot{p}_i\vect{D}_k\big([\vect{R}_{il}]_{L}^\mathrm{B}
-\vect{C}_{\vect{\hat{h}}_i}\big)\vect{D}_k$ \cite{Nayebi2016Performance}.
Motivated by this observation and by replacing $\vect{\hat{h}}_i$ with $\vect{D}_i\vect{\hat{h}}_i$, we alternatively propose the P-MMSE vector, $\vect{v}_{k}^\mathrm{P-MMSE}$, as follows
\begin{align}\label{P-MMSE_detector}
\left\{
\begin{array}{lr}
\vect{v}_{k}^\mathrm{P-MMSE}=(\vect{B}_k^\mathrm{P})^\dagger\vect{D}_k\vect{\hat{h}}_k,
\\\vect{B}_k^\mathrm{P}=\vect{D}_k\bigg\{
(1-\rho_\mathrm{ad})^2\sum_{i\in\big(\mathcal{Q}_k-\mathcal{N}_{l_k^\mathrm{M}}\big)}\ddot{p}_i\big(\vect{\bar{h}}_i\vect{\bar{h}}_i^\mathrm{H}+[\vect{R}_{il}]_{L}^\mathrm{B}\big)
\\+(1-\rho_\mathrm{ad})^2\sum_{i\in\mathcal{Q}_k\cap\mathcal{N}_{l_k^\mathrm{M}}}\ddot{p}_i(\vect{\hat{h}}_i\vect{\hat{h}}_i^\mathrm{H}+[\vect{R}_{il}]_{L}^\mathrm{B}
-\vect{C}_{\vect{\hat{h}}_i})
\\+(1-\rho_\mathrm{ad})\sigma^2\vect{I}_{NL}
+\bigg[\frac{(1-\rho_\mathrm{ad})^2\rho_\mathrm{da}}{1-\rho_\mathrm{da}}\sum_{i\in\mathcal{Q}_k}{\ddot{p}_i}\big(\vect{\bar{h}}_{il}\vect{\bar{h}}_{il}^\mathrm{H}+\vect{R}_{il}\big)
\\+\frac{\rho_\mathrm{ad}(1-\rho_\mathrm{ad})}{1-\rho_\mathrm{da}}\diag\bigg(\sum_{i\in\mathcal{Q}_k}{\ddot{p}_i}\big(\vect{\bar{h}}_{il}\vect{\bar{h}}_{il}^\mathrm{H}+\vect{R}_{il}\big)\bigg)
\bigg]_{L}^\mathrm{B}\bigg\}\vect{D}_k,
\end{array}
\right.
\end{align}
where $\mathcal{Q}_k=\{i|\vect{D}_k\vect{D}_i\neq\vect{0}_{LN},i=1,2,\cdots,K\}$, and $\mathcal{N}_{l_k^\mathrm{M}}$ is the UE set served by the P-AP of $k^\mathrm{th}$ UE. As was the case with \eqref{LP-MMSE_detector}, this new vector has lower CC while maintaining excellent SE performance as compared to the original P-MMSE in \cite{Emil2020scalable} (see Fig. \ref{UL-SE_CDF}(b) for details).

\section{Spectral Efficiency Analysis with MRC Detection}\label{section:Spectral Efficiency Analysis}
Although MMSE combining yields optimal SE performance, its CC for combining vectors is high and might be even prohibitive to be used for IoT applications. Alternatively, the MRC detection can be used as it has relatively low complexity while maximizing the signal-to-noise ratio (SNR) of the desired signal. In this section we will analyze the SE performance for our proposed SCF-mMIMO system using MRC detection.

\subsection{SE Analysis for Distributed Scheme}
To obtain the desired analytical SE expression, first Theorem \ref{Theorem-1} is introduced, followed by Theorem \ref{Theorem-2} where a closed-form SE expression will be presented.
\begin{theorem}\label{Theorem-1}
The term $\mathbb{E}[\vect{\hat{h}}_{kl_1}^\mathrm{H}\vect{{h}}_{il_1}\vect{{h}}_{il_2}^\mathrm{H}\vect{\hat{h}}_{kl_2}]$ can be evaluated as \eqref{expectation1-a-1}--\eqref{expectation1-a-4}.
\end{theorem}
\begin{figure*}[!t]
\begin{align}
\notag&\mathbb{E}[\vect{\hat{h}}_{kl_1}^\mathrm{H}\vect{h}_{il_1}\vect{h}_{il_2}^\mathrm{H}\vect{\hat{h}}_{kl_2}]|_{(i\in\mathcal{P}_k ,l_1=l_2)}=|\vect{\bar{h}}_{kl_1}^\mathrm{H}\vect{\bar{h}}_{il_1}|^2
+(1-\rho_\mathrm{ad})^2\tau\ddot{p}_k\big[\tr\big(\vect{R}_{il_1}\vect{R}_{kl_1}\vect{\Psi}_{t_k^\mathrm{p},l_1}^{-1}\vect{R}_{kl_1}\big)
+\vect{\bar{h}}_{il_1}^\mathrm{H}\vect{R}_{kl_1}\vect{\Psi}_{t_k^\mathrm{p},l_1}^{-1}\vect{R}_{kl_1}\vect{\bar{h}}_{il_1}\big]
\\&+\vect{\bar{h}}_{kl_1}^\mathrm{H}\vect{R}_{il_1}\vect{\bar{h}}_{kl_1}
+(1-\rho_\mathrm{ad})^4\tau^2\ddot{p}_k\ddot{p}_i|\tr(\vect{R}_{il_1}\vect{\Psi}_{t_k^\mathrm{p},l_1}^{-1}\vect{R}_{kl_1})|^2
+2(1-\rho_\mathrm{ad})^2\tau\sqrt{\ddot{p}_i\ddot{p}_k}\Re\big\{\tr\big(\vect{R}_{il_1}\vect{\Psi}_{t_k^\mathrm{p},l_1}^{-1}\vect{R}_{kl_1}\big)\vect{\bar{h}}_{il_1}^\mathrm{H}\vect{\bar{h}}_{kl_1}\big\}
\label{expectation1-a-1}
\\\notag&\mathbb{E}[\vect{\hat{h}}_{kl_1}^\mathrm{H}\vect{h}_{il_1}\vect{h}_{il_2}^\mathrm{H}\vect{\hat{h}}_{kl_2}]|_{(i\in\mathcal{P}_k ,l_1\neq l_2)}=\vect{\bar{h}}_{kl_1}^\mathrm{H}\vect{\bar{h}}_{il_1}\vect{\bar{h}}_{il_2}^\mathrm{H}\vect{\bar{h}}_{kl_2}
+(1-\rho_\mathrm{ad})^4\tau^2
\ddot{p}_k\ddot{p}_i\tr(\vect{R}_{il_1}\vect{\Psi}_{t_k^\mathrm{p},l_1}^{-1}\vect{R}_{kl_1})\tr(\vect{R}_{kl_2}\vect{\Psi}_{t_k^\mathrm{p},l_2}^{-1}\vect{R}_{il_2})
\\&+(1-\rho_\mathrm{ad})^2\tau\sqrt{\ddot{p}_k\ddot{p}_i}\big[\vect{\bar{h}}_{il_2}^\mathrm{H}\vect{\bar{h}}_{kl_2}\tr\big(\vect{R}_{il_1}\vect{\Psi}_{t_k^\mathrm{p},l_1}^{-1}\vect{R}_{kl_1}\big)
+\vect{\bar{h}}_{kl_1}^\mathrm{H}\vect{\bar{h}}_{il_1}\tr\big(\vect{R}_{kl_2}\vect{\Psi}_{t_k^\mathrm{p},l_2}^{-1}\vect{R}_{il_2}\big)\big]
\label{expectation1-a-2}
\\\notag&\mathbb{E}[\vect{\hat{h}}_{kl_1}^\mathrm{H}\vect{h}_{il_1}\vect{h}_{il_2}^\mathrm{H}\vect{\hat{h}}_{kl_2}]|_{(i\notin\mathcal{P}_k ,l_1=l_2)}=|\vect{\bar{h}}_{kl_1}^\mathrm{H}\vect{\bar{h}}_{il_1}|^2
+\vect{\bar{h}}_{kl_1}^\mathrm{H}\vect{R}_{il_1}\vect{\bar{h}}_{kl_1}
\\&+(1-\rho_\mathrm{ad})^2\tau\ddot{p}_k\big[\tr\big(\vect{R}_{il_1}\vect{R}_{kl_1}\vect{\Psi}_{t_k^\mathrm{p},l_1}^{-1}\vect{R}_{kl_1}\big)
+\vect{\bar{h}}_{il_1}^\mathrm{H}\vect{R}_{kl_1}\vect{\Psi}_{t_k^\mathrm{p},l_1}^{-1}\vect{R}_{kl_1}\vect{\bar{h}}_{il_1}\big]
\label{expectation1-a-3}
\\&\mathbb{E}[\vect{\hat{h}}_{kl_1}^\mathrm{H}\vect{h}_{il_1}\vect{h}_{il_2}^\mathrm{H}\vect{\hat{h}}_{kl_2}]|_{(i\notin\mathcal{P}_k ,l_1\neq l_2)}=\vect{\bar{h}}_{kl_1}^\mathrm{H}\vect{\bar{h}}_{il_1}\vect{\bar{h}}_{il_2}^\mathrm{H}\vect{\bar{h}}_{kl_2}
\label{expectation1-a-4}
\end{align}
\hrulefill
\end{figure*}
\begin{IEEEproof}
See Appendix \ref{Appendix:Theorem-1}.
\end{IEEEproof}
\begin{theorem}\label{Theorem-2}
Assuming that MRC detection is employed by all APs and that LSFD is used by the CPU, and by using Theorem \ref{Theorem-1}, the maximum SE of $k^\mathrm{th}$ UE in \eqref{UL-SE-max} can be obtained in closed-form as follows
\begin{align}\label{UL-SE-MRC}
&\notag{R}_{k,\max}^\mathrm{d}=\Big(1-\frac{\tau}{\tau_c}\Big)\times
\\&\log_2\Big(1+(1-\rho_\mathrm{ad})^2\ddot{p}_k(\bm{\lambda}_{k}^k+\vect{b}_k^k)^\mathrm{H}\vect{C}_k^{-1}(\bm{\lambda}_{k}^k+\vect{b}_k^k)\Big)
\end{align}
where $\vect{b}_k^i=[b_{kl}^i|l\in\mathcal{M}_k]^\mathrm{T}$, $\bm{\lambda}_{k}^i=[\lambda_{kl}^i|l\in\mathcal{M}_k]^\mathrm{T}$ with
$b_{kl}^i=(1-\rho_\mathrm{ad})^2\tau\sqrt{\ddot{p}_k\ddot{p}_i}\tr\big(\vect{R}_{il}\vect{\Psi}_{t_k^\mathrm{p},l}^{-1}\vect{R}_{kl}\big),$
$\lambda_{kl}^i=\vect{\bar{h}}_{kl}^\mathrm{H}\vect{\bar{h}}_{il}$ and
\begin{align}\label{C_k}
\notag\vect{C}_k&=
\frac{(1-\rho_\mathrm{ad})^2}{1-\rho_\mathrm{da}}\Bigg\{\sum_{i=1}^{K}\ddot{p}_i\Big[\bm{\lambda}_{k}^i(\bm{\lambda}_{k}^i)^\mathrm{H}
+\big[c_{kl}^i\big]_{|\mathcal{M}_k|}^\Lambda\Big]
\\\notag&+\sum_{i\in\mathcal{P}_k}\ddot{p}_i\Big[\vect{b}_k^i(\vect{b}_k^i)^\mathrm{H}
+\vect{b}_k^i(\bm{\lambda}_{k}^i)^\mathrm{H}+\bm{\lambda}_{k}^i(\vect{b}_k^i)^\mathrm{H}\Big]\Bigg\}
\\&-(1-\rho_\mathrm{ad})^2\ddot{p}_k(\bm{\lambda}_{k}^k+\vect{b}_k^k)(\bm{\lambda}_{k}^k+\vect{b}_k^k)^\mathrm{H}
+\big[d_{kl}\big]_{|\mathcal{M}_k|}^\Lambda
\end{align}
with $d_{kl}$ as \eqref{d_kl}
\begin{figure*}[!t]
\begin{align}\notag\label{d_kl}
d_{kl}&=\frac{\rho_\mathrm{ad}(1-\rho_\mathrm{ad})}{1-\rho_\mathrm{da}}\vect{\bar{h}}_{kl}^\mathrm{H}\diag\bigg(\sum_{i=1}^{K}{\ddot{p}_i}\vect{R}_{il}\bigg)\vect{\bar{h}}_{kl}
+\frac{\rho_\mathrm{ad}(1-\rho_\mathrm{ad})^3}{1-\rho_\mathrm{da}}\tau\ddot{p}_k\tr\bigg(\diag\bigg(\sum_{i=1}^{K}\ddot{p}_i\vect{R}_{il}\bigg)\vect{R}_{kl}\vect{\Psi}_{t_k^\mathrm{p},l}^{-1}\vect{R}_{kl}\bigg)
\\&\quad+(1-\rho_\mathrm{ad})\bigg[\sigma^2+\frac{\rho_\mathrm{ad}}{1-\rho_\mathrm{da}}\sum_{i=1}^{K}\frac{\ddot{p}_i\beta_{il}\kappa_{il}}{\kappa_{il}+1}\bigg]
\big[\vect{\bar{h}}_{kl}^\mathrm{H}\vect{\bar{h}}_{kl}+(1-\rho_\mathrm{ad})^2\tau\ddot{p}_k\tr\big(\vect{R}_{kl}\vect{\Psi}_{t_k^\mathrm{p},l}^{-1}\vect{R}_{kl}\big)\big]
\end{align}
\hrulefill
\end{figure*}
and
\begin{align}
\notag{c}_{kl}^i&=(1-\rho_\mathrm{ad})^2\tau\ddot{p}_k\tr\big(\vect{R}_{il}\vect{R}_{kl}\vect{\Psi}_{t_k^\mathrm{p},l}^{-1}\vect{R}_{kl}\big)
+\vect{\bar{h}}_{kl}^\mathrm{H}\vect{R}_{il}\vect{\bar{h}}_{kl}
\\&\quad+(1-\rho_\mathrm{ad})^2\tau\ddot{p}_k\vect{\bar{h}}_{il}^\mathrm{H}\vect{R}_{kl}\vect{\Psi}_{t_k^\mathrm{p},l}^{-1}\vect{R}_{kl}\vect{\bar{h}}_{il}.
\end{align}
\end{theorem}
\begin{IEEEproof}
See Appendix \ref{Appendix:Theorem-2}.
\end{IEEEproof}
From \eqref{ex-Th1-1} and \eqref{ex-Th1-2}, one can obtain that the terms $\mathbb{E}[\vect{f}_k\vect{f}_k^\mathrm{H}]$ and $\sum_{i=1}^{K}\ddot{p}_i\mathbb{E}[\vect{g}_{ki}\vect{g}_{ki}^\mathrm{H}]$ are related as
\begin{equation}
\mathbb{E}[\vect{f}_k\vect{f}_k^\mathrm{H}]=\big[d_{kl}\big]_{|\mathcal{M}_k|}^\Lambda
+\frac{(1-\rho_\mathrm{ad})^2\rho_\mathrm{da}}{1-\rho_\mathrm{da}}\sum_{i=1}^{K}\ddot{p}_i\mathbb{E}[\vect{g}_{ki}\vect{g}_{ki}^\mathrm{H}].
\end{equation}
Using the above result, $\vect{a}_k$ in \eqref{LSFD} can be rewritten as
\begin{align}\label{LSFD-2}
\notag\vect{a}_k^\mathrm{MR}&=\bigg[\big[d_{kl}\big]_{|\mathcal{M}_k|}^\Lambda
+\frac{(1-\rho_\mathrm{ad})^2}{1-\rho_\mathrm{da}}\sum_{i=1}^{K}\ddot{p}_i\mathbb{E}[\vect{g}_{ki}\vect{g}_{ki}^\mathrm{H}]
\\&\quad-(1-\rho_\mathrm{ad})^2\ddot{p}_k\mathbb{E}[\vect{g}_{kk}]\mathbb{E}[\vect{g}_{kk}^\mathrm{H}]\bigg]^{-1}\mathbb{E}[\vect{g}_{kk}]
\end{align}
which is unscalable because the number of complex multiplications required to calculate
$\sum_{i=1}^{K}\ddot{p}_i\mathbb{E}[\vect{g}_{ki}\vect{g}_{ki}^\mathrm{H}]$ in \eqref{LSFD-2} grows arbitrarily with $K$.
Similarly to the previously discussed centralized scheme (see \eqref{P-MMSE_detector}), here we propose the P-LSFD vector, $\vect{a}_k^\mathrm{P-MR}$, as follows
\begin{align}\label{P-LSFD}
\notag&\vect{a}_k^\mathrm{P-MR}=\bigg[\big[d_{kl}\big]_{|\mathcal{M}_k|}^\Lambda
+\frac{(1-\rho_\mathrm{ad})^2}{1-\rho_\mathrm{da}}\sum_{i\in\mathcal{Q}_k}\ddot{p}_i\mathbb{E}[\vect{g}_{ki}\vect{g}_{ki}^\mathrm{H}]
\\\notag&\qquad\qquad\;-(1-\rho_\mathrm{ad})^2\ddot{p}_k\mathbb{E}[\vect{g}_{kk}]\mathbb{E}[\vect{g}_{kk}^\mathrm{H}]\bigg]^{-1}\mathbb{E}[\vect{g}_{kk}]
\\&\qquad\quad=\big(\vect{C}_k^\mathrm{P}\big)^{-1}(\bm{\lambda}_{k}^k+\vect{b}_k^k),
\\\notag&\vect{C}_k^\mathrm{P}=
\frac{(1-\rho_\mathrm{ad})^2}{1-\rho_\mathrm{da}}\Bigg\{\sum_{i\in\mathcal{Q}_k}\ddot{p}_i\Big[\bm{\lambda}_{k}^i(\bm{\lambda}_{k}^i)^\mathrm{H}
+\big[c_{kl}^i\big]_{|\mathcal{M}_k|}^\Lambda\Big]
\\\notag&+\sum_{i\in\mathcal{P}_k\cap\mathcal{Q}_k}\ddot{p}_i\Big[\vect{b}_k^i(\vect{b}_k^i)^\mathrm{H}
+\vect{b}_k^i(\bm{\lambda}_{k}^i)^\mathrm{H}+\bm{\lambda}_{k}^i(\vect{b}_k^i)^\mathrm{H}\Big]\Bigg\}
\\&-(1-\rho_\mathrm{ad})^2\ddot{p}_k(\bm{\lambda}_{k}^k+\vect{b}_k^k)(\bm{\lambda}_{k}^k+\vect{b}_k^k)^\mathrm{H}
+\big[d_{kl}\big]_{|\mathcal{M}_k|}^\Lambda.
\end{align}
It is noted that $\vect{a}_k^\mathrm{P-MR}$ becomes identical to $\vect{a}_k^\mathrm{MR}$ when $k^\mathrm{th}$ UE is served by all APs.

In order to derive an analytical expression for the SE, the following corollary, obtained from Theorem \ref{Theorem-2}, will be used.
\begin{corollary}
With MRC detection used by each AP and P-LSFD used by CPU, the UL-SE of $k^\mathrm{th}$ UE of \eqref{UL-SE} is given in the closed form as
\begin{align}\label{UL-SE-MRC-PLSFD}
\notag&{R}_{k}^\mathrm{d}=\Big(1-\frac{\tau}{\tau_c}\Big)\times
\\&\log_2\bigg(1+\frac{(1-\rho_\mathrm{ad})^2\ddot{p}_k\big|(\vect{a}_k^\mathrm{P-MR})^\mathrm{H}(\bm{\lambda}_{k}^k+\vect{b}_k^k)\big|^2}
{(\vect{a}_k^\mathrm{P-MR})^\mathrm{H}\vect{C}_k\vect{a}_k^\mathrm{P-MR}}\bigg).
\end{align}
\end{corollary}
\begin{IEEEproof}
From \eqref{UL-SE} and by using its SINR term (i.e., \eqref{SNR-d}) and by replacing $\vect{a}_k$, $\mathbb{E}[\vect{g}_{kk}]$, and $\vect{B}_k^\mathrm{d}$ with $\vect{a}_k^\mathrm{P-MR}$, $\bm{\lambda}_{k}^k+\vect{b}_k^k$, and $\vect{C}_k$, respectively, \eqref{UL-SE-MRC-PLSFD} can be readily obtained.
\end{IEEEproof}

Based on \eqref{P-LSFD}, the CC of our proposed P-LSFD vector will be derived and compared with the conventional unscalable methods, which is shown as the following proposition.
\begin{proposition}\label{prop:complexity_for_MMSE}
The CC for calculating the P-LSFD vector in \eqref{P-LSFD} requires $\big[|\mathcal{M}_k|(|\mathcal{M}_k|+1)|\mathcal{Q}_k|/2+|\mathcal{M}_k|(5|\mathcal{M}_k|+1)|\mathcal{P}_k\cap\mathcal{Q}_k|/2
+(|\mathcal{M}_k|^3+3|\mathcal{M}_k|^2-|\mathcal{M}_k|)/3\big]$ complex multiplications (CMs) and $|\mathcal{M}_k|$ complex divisions (CDs).
\end{proposition}
\begin{IEEEproof}
Since $\vect{b}_k^i,\bm{\lambda}_{k}^i\in\mathbb{C}^{|\mathcal{M}_k|\times1}$, by using \cite[Lemma B.$1$]{Emil2017mMnetworks}, both $\bm{\lambda}_{k}^i(\bm{\lambda}_{k}^i)^\mathrm{H}$ and $\vect{b}_k^i(\vect{b}_k^i)^\mathrm{H}$ can be obtained by using $|\mathcal{M}_k|(|\mathcal{M}_k|+1)/2$ CMs based on their Hermitian symmetry. In addition, both $\bm{\lambda}_{k}^i(\vect{b}_k^i)^\mathrm{H}$ and $\vect{b}_k^i(\bm{\lambda}_{k}^i)^\mathrm{H}$ can be obtained by using $|\mathcal{M}_k|^2$ CMs.
Thus $\vect{C}_k^\mathrm{P}$ can be obtained by using $\big[|\mathcal{M}_k|(|\mathcal{M}_k|+1)|\mathcal{Q}_k|/2+|\mathcal{M}_k|(5|\mathcal{M}_k|+1)|\mathcal{P}_k\cap\mathcal{Q}_k|/2$ CMs.
Since $\vect{C}_k^\mathrm{P}\in\mathbb{C}^{|\mathcal{M}_k|\times|\mathcal{M}_k|}$, by using \cite[Lemma B.$2$]{Emil2017mMnetworks}, the $\vect{L}\vect{D}\vect{L}^\mathrm{H}$ decomposition of $\vect{C}_k^\mathrm{P}$
requires $(|\mathcal{M}_k|^3-|\mathcal{M}_k|)/3$ CMs. Finally, $\vect{a}_k^\mathrm{P-MR}=\big(\vect{C}_k^\mathrm{P}\big)^{-1}(\bm{\lambda}_{k}^k+\vect{b}_k^k)$ can be obtained by using $|\mathcal{M}_k|^2$ CMs and $|\mathcal{M}_k|$ CDs. Q.E.D.
\end{IEEEproof}

By using the same method as Proposition \ref{prop:complexity_for_MMSE}, the CC for other weighting vectors can be found in Table \ref{tab:cost weighting vector}.
It is noted that, due to the AP cluster formation adopted by each UE, the CC for all weighting vector schemes are independent of $L$, which is an inherent advantage of the SCF-mMIMO operation.
Furthermore, it can also be observed that, contrary to the conventional unscalable LSFD vector in \cite{Emil2020scalable}, the CC of our proposed P-LSFD vector scheme are also independent of $K$ and this achieves the scalability at the CC level.
\begin{table*}[t]
\renewcommand{\arraystretch}{1.}
\caption{\textsc{CC for $k^\mathrm{th}$ UE with Different Weighting Vector for Each Realization of the UEs' Locations}}
\vskip 1mm
\label{tab:cost weighting vector}
\centering
\begin{tabular}{|c|c|}
\hline
Scheme&Computational Complexity\\
\hline
P-LSFD&$\left[|\mathcal{M}_k|(|\mathcal{M}_k|+1)|\mathcal{Q}_k|/2+|\mathcal{M}_k|(5|\mathcal{M}_k|+1)|\mathcal{P}_k\cap\mathcal{Q}_k|/2
+(|\mathcal{M}_k|^3+3|\mathcal{M}_k|^2-|\mathcal{M}_k|)/3\right]$ CMs and $|\mathcal{M}_k|$ CDs\\
\hline
LSFD\cite{Emil2020scalable}&$\left[|\mathcal{M}_k|(|\mathcal{M}_k|+1)K/2+|\mathcal{M}_k|(5|\mathcal{M}_k|+1)|\mathcal{P}_k|/2
+(|\mathcal{M}_k|^3+3|\mathcal{M}_k|^2-|\mathcal{M}_k|)/3\right]$ CMs and $|\mathcal{M}_k|$ CDs\\
\hline
L2-LSFD\cite{Emil2020scalable}&Null\\
\hline
\end{tabular}
\end{table*}

\subsection{SE Analysis for Centralized Scheme}
\begin{theorem}\label{Theorem-3}
With MRC detection used by CPU, the SE of $k^\mathrm{th}$ UE in \eqref{UL-SE-C} can be obtained in the closed form as \eqref{UL-SE-C-approx-close},
\begin{figure*}[!t]
\begin{equation}\label{UL-SE-C-approx-close}
\widehat{R_{k}^\mathrm{c}}=\Big(1-\frac{\tau}{\tau_c}\Big)\log_2\Bigg(1+\frac{(1-\rho_\mathrm{ad})^2\ddot{p}_k[f_k^\mathrm{g}(k)+f_k^\mathrm{e}(k)]}
{(1-\rho_\mathrm{ad})^2\big[\sum_{i\neq k}^{K}\ddot{p}_if_k^\mathrm{g}(i)+\sum_{i\in\mathcal{P}_k\setminus{k}}^{K}\ddot{p}_if_k^\mathrm{e}(i)\big]
+\tr\big(\vect{D}_k^\mathrm{c}\big(\vect{\bar{h}}_{k}\vect{\bar{h}}_{k}^\mathrm{H}+\vect{C}_{\vect{\hat{h}}_k}\big)\big)}\Bigg)
\end{equation}
\hrulefill
\end{figure*}
where
$\vect{D}_k^\mathrm{c}=\vect{D}_k\big[(1-\rho_\mathrm{ad})^2\sum_{i=1}^{K}\ddot{p}_i\big([\vect{R}_{il}]_{L}^\mathrm{B}
-\vect{C}_{\vect{\hat{h}}_i}\big)+[\vect{C}_{\mathrm{n},l}]_{L}^\mathrm{B}\big]$ and
\begin{align}
\notag &f_k^\mathrm{g}(i)=(1-\rho_\mathrm{ad})^4\tau^2\ddot{p}_k\ddot{p}_i\tr([\vect{D}_{kl}\vect{R}_{kl}\vect{\Psi}_{t_k^\mathrm{p},l}^{-1}\vect{R}_{kl}\vect{R}_{il}\vect{\Psi}_{t_i^\mathrm{p},l}^{-1}\vect{R}_{il}]_L^\mathrm{B})
\\\notag&+|\vect{\bar{h}}_{k}^\mathrm{H}\vect{D}_k\vect{\bar{h}}_{i}|^2
+(1-\rho_\mathrm{ad})^2\tau\ddot{p}_i\vect{\bar{h}}_{k}^\mathrm{H}[\vect{D}_{kl}\vect{R}_{il}\vect{\Psi}_{t_i^\mathrm{p},l}^{-1}\vect{R}_{il}]_L^\mathrm{B}\vect{\bar{h}}_{k}
\\&+(1-\rho_\mathrm{ad})^2\tau\ddot{p}_k\vect{\bar{h}}_{i}^\mathrm{H}[\vect{D}_{kl}\vect{R}_{kl}\vect{\Psi}_{t_k^\mathrm{p},l}^{-1}\vect{R}_{kl}]_L^\mathrm{B}\vect{\bar{h}}_{i},
\\\notag&{f}_k^\mathrm{e}(i)=(1-\rho_\mathrm{ad})^4\tau^2\ddot{p}_k\ddot{p}_i|\tr([\vect{D}_{kl}\vect{R}_{il}\vect{\Psi}_{t_k^\mathrm{p},l}^{-1}\vect{R}_{kl}]_L^\mathrm{B})|^2
\\&+2(1-\rho_\mathrm{ad})^2\tau\sqrt{\ddot{p}_i\ddot{p}_k}\Re\big\{\tr\big([\vect{D}_{kl}\vect{R}_{il}\vect{\Psi}_{t_k^\mathrm{p},l}^{-1}\vect{R}_{kl}]_L^\mathrm{B}\big)\vect{\bar{h}}_{i}^\mathrm{H}\vect{D}_k\vect{\bar{h}}_{k}\big\}.
\end{align}
\end{theorem}
\begin{IEEEproof}
Applying \cite[Lemma $1$]{Q.Zhang2014a}, $R_{k}^\mathrm{c}$ in \eqref{UL-SE-C} can be approximated by $\widehat{R_{k}^\mathrm{c}}$ as \eqref{UL-SE-C-approx}.
\begin{figure*}[!t]
\begin{equation}\label{UL-SE-C-approx}
\widehat{R_{k}^\mathrm{c}}=\Big(1-\frac{\tau}{\tau_c}\Big)\log_2\Bigg(1+
\frac{(1-\rho_\mathrm{ad})^2\ddot{p}_k\mathbb{E}[|\vect{\hat{h}}_{k}^\mathrm{H}\vect{D}_k\vect{\hat{h}}_k|^2]}
{(1-\rho_\mathrm{ad})^2\sum_{i\neq k}^{K}\ddot{p}_i\mathbb{E}[|\vect{\hat{h}}_{k}^\mathrm{H}\vect{D}_k\vect{\hat{h}}_i|^2]
+\mathbb{E}\big[\vect{\hat{h}}_{k}^\mathrm{H}\vect{D}_k^\mathrm{c}\vect{D}_k\vect{\hat{h}}_{k}\big]}\Bigg)
\end{equation}
\hrulefill
\end{figure*}
Since $\vect{z}_{t_k^\mathrm{p},l}^\mathrm{w}=\vect{z}_{t_i^\mathrm{p},l}^\mathrm{w}$ for $\forall{i}\in\mathcal{P}_k$, and following \eqref{central_h_hat} and \eqref{h-hat-sim}, $\vect{\hat{h}}_{k}$ and $\vect{\hat{h}}_{i}$ can be expressed as
\begin{align}
\vect{\hat{h}}_{k}&=\vect{\bar{h}}_{k}+(1-\rho_\mathrm{ad})\sqrt{\ddot{p}_k\tau}\big[\vect{R}_{kl}\vect{\Psi}_{t_k^\mathrm{p},l}^{-\frac{1}{2}}\big]_{L}^\mathrm{B}\vect{w},
\\\vect{\hat{h}}_{i}&=\vect{\bar{h}}_{i}+(1-\rho_\mathrm{ad})\sqrt{\ddot{p}_i\tau}\big[\vect{R}_{il}\vect{\Psi}_{t_i^\mathrm{p},l}^{-\frac{1}{2}}\big]_{L}^\mathrm{B}\vect{w},
\end{align}
where $\vect{w}\sim\mathcal{CN}(\vect{0},\vect{I}_{LN})$ and $\vect{\Psi}_{t_k^\mathrm{p},l}=\vect{\Psi}_{t_i^\mathrm{p},l}$.
For $\vect{h}_1=\vect{D}_k\vect{\hat{h}}_{k}$ and $\vect{h}_2=\vect{\hat{h}}_{i}$ using \cite[Lemma 5]{Ozdogan2019Massive}, it can be easily obtained that $\mathbb{E}[|\vect{\hat{h}}_{k}^\mathrm{H}\vect{D}_k\vect{\hat{h}}_{i}|^2]|_{i\in\mathcal{P}_k}=f_k^\mathrm{g}(i)+f_k^\mathrm{e}(i)$.

Since $\vect{z}_{t_k,l}^\mathrm{w}$ and $\vect{z}_{t_i,l}^\mathrm{w}$ are independent for $\forall{i}\notin\mathcal{P}_k$, $\vect{\hat{h}}_{k}$ and $\vect{\hat{h}}_{i}$ can be reformulated as
\begin{align}
\vect{\hat{h}}_{k}&=\vect{\bar{h}}_{k}+(1-\rho_\mathrm{ad})\sqrt{\ddot{p}_k\tau}\big[\vect{R}_{kl}\vect{\Psi}_{t_k^\mathrm{p},l}^{-\frac{1}{2}}\big]_{L}^\mathrm{B}\vect{w}_k,
\\\vect{\hat{h}}_{i}&=\vect{\bar{h}}_{i}+(1-\rho_\mathrm{ad})\sqrt{\ddot{p}_i\tau}\big[\vect{R}_{il}\vect{\Psi}_{t_i^\mathrm{p},l}^{-\frac{1}{2}}\big]_{L}^\mathrm{B}\vect{w}_i,
\end{align}
where $\vect{w}_k\sim\mathcal{CN}(\vect{0},\vect{I}_{LN})$ and $\vect{w}_i\sim\mathcal{CN}(\vect{0},\vect{I}_{LN})$ with $\mathrm{E}[\vect{w}_k\vect{w}_i^\mathrm{H}]=\vect{0}$. Furthermore, by again substituting $\vect{h}_1=\vect{D}_k\vect{\hat{h}}_{k}$ and $\vect{h}_2=\vect{\hat{h}}_{i}$ into \cite[Lemma 4]{Ozdogan2019Massive}, one can obtain that $\mathbb{E}[|\vect{\hat{h}}_{k}^\mathrm{H}\vect{D}_k\vect{\hat{h}}_{i}|^2]|_{i\notin\mathcal{P}_k}=f_k^\mathrm{g}(i)$.
This completes the proof since $\mathbb{E}\big[\vect{\hat{h}}_{k}^\mathrm{H}\vect{D}_k^\mathrm{c}\vect{D}_k\vect{\hat{h}}_{k}\big]=\tr\big(\vect{D}_k^\mathrm{c}\big(\vect{\bar{h}}_{k}\vect{\bar{h}}_{k}^\mathrm{H}+\vect{C}_{\vect{\hat{h}}_k}\big)\big)$.
\end{IEEEproof}

\subsection{CC Comparison between Different MMSE Detectors}\label{cc}
\begin{proposition}
Our proposed $\vect{v}_{k}^\mathrm{LP-MMSE}$ and $\vect{v}_{k}^\mathrm{P-MMSE}$ detectors presented in \eqref{LP-MMSE_detector} and \eqref{P-MMSE_detector} respectively, have $|\mathcal{N}_l^\mathrm{S}|$ and $|\mathcal{Q}_k-\mathcal{N}_{l_k^\mathrm{M}}|$ less CMs than the original scalable MMSE detectors in \cite{Emil2020scalable}, respectively.
\end{proposition}
\begin{IEEEproof}
In \eqref{zp}, since $\vect{Z}_l\in\mathbb{C}^{N\times\tau}$ and $\bm{\phi}_{t_k^\mathrm{p}}^\ast\in\mathbb{C}^{\tau\times1}$, $\vect{z}_{t_k^\mathrm{p},l}$ can be obtained by using $N\tau$ CMs. In \eqref{hhat}, since $\vect{R}_{kl}\vect{\Psi}_{t_k^\mathrm{p},l}^{-1}\in\mathbb{C}^{N\times{N}}$ and $\vect{z}_{t_k^\mathrm{p},l}\in\mathbb{C}^{N\times1}$, $\vect{\hat{h}}_{kl}$ can be obtained by using $N^2$ CMs. In order to calculate $\vect{v}_{k}^\mathrm{LP-MMSE}$ in \eqref{LP-MMSE_detector}, the set of estimated channel $\{\vect{\hat{h}}_{il}|i\in\mathcal{N}_l^\mathrm{P}\}$ are needed, thus the total CMs required for CE to get $\vect{v}_{k}^\mathrm{LP-MMSE}$ in \eqref{LP-MMSE_detector} becomes $N(N+\tau)|\mathcal{N}_l^\mathrm{P}|$.
By replacing $\mathcal{N}_l^\mathrm{P}$ with $\mathcal{N}_l$ , the total CMs required for CE to get $\vect{v}_{k}^\mathrm{LP-MMSE}$ in \cite[(29)]{Emil2020scalable} is $N(N+\tau)|\mathcal{N}_l|$. Likewise, the total CMs required for CE to get $\vect{v}_{k}^\mathrm{P-MMSE}$ in \eqref{P-MMSE_detector} and \cite[(23)]{Emil2020scalable} are $N(N+\tau)|\mathcal{Q}_k\cap\mathcal{N}_{l_k^\mathrm{M}}||\mathcal{M}_k|$ and $N(N+\tau)|\mathcal{Q}_k||\mathcal{M}_k|$, respectively. Q.E.D.
\end{IEEEproof}

\section{Scalable Algorithm for Joint AP Cluster Formation, Pilot Assignment, and Power Control}
\label{section:algorithm}
For each accessing UE, firstly it must be assigned to a P-AP with the largest value of large-scale fading to avoid the case in which it is not served by any AP. After that, other S-APs will decide whether to serve it. This whole procedure is called the AP cluster formation.
The previous AP cluster formation schemes in \cite{Emil2020scalable} and \cite{Chen2020Structured} confines the number of UEs served by each AP (including the P-AP) not to excess the pilot length, which leads to the frequent access competition for the same AP.
Since $LN\gg{K}$ and each UE is uniformly distributed \cite{Emil2020scalable}, $K\rightarrow\infty$ requires also $L\rightarrow\infty$. Thus, the number of UEs who appoint the same AP as their P-AP is finite as $K\rightarrow\infty$;
Therefore, in our algorithm, we have removed the limitation on the number of accessing UE for the same P-AP but only confine the number of UEs served by the same S-AP.
Comparing with the pilot assignment scheme proposed in \cite{Chen2020Structured}, our pilot assignment for each UE is carried out simultaneously with the assignment of its P-AP, which not only yields better SE performance but also maintains much less CC.
Furthermore, in order to achieve the required UL QoS fairness among each accessing UE, we adopt the fractional power control first proposed in \cite{Nikbakht2020}, where $\ddot{p}_k$ is set proportionally to $(\sum_{l\in\mathcal{M}_k}\beta_{kl})^{-1}$ to minimize the variance of the SIR. The reason for this choice is the fact that a smaller variance translates to a smaller fluctuation around the mean value for each specific UE's SIR, which will eventually guarantee the QoS fairness of each UE.

The proposed algorithm (Algorithm \ref{alg1}) can be implemented using the following steps:
\begin{enumerate}
  \item[\textit{i})] During the P-AP assignment, firstly the $k^\mathrm{th}$ UE obtains the large scale faded signal powers, $\beta_{kl}$, received from the set $\mathcal{L}(\bar{d})\subsetneqq\{1,2,\cdots,L\}$. Then it selects AP ${l_k^\mathrm{M}}$ with the largest $\beta_{kl}$ to act as its P-AP, i.e., $l_k^\mathrm{M}=\arg\max_{l\in\mathcal{L}(\bar{d})}\beta_{kl}$.
  \item[\textit{ii})] From \eqref{phi_tp}, for $k>\tau$, the selected P-AP assigns $t_k^\mathrm{p}$ to the $k^\mathrm{th}$ UE with
      \begin{equation}
      t_k^\mathrm{p}=\arg\min_{t^\mathrm{p}\in\{1,2,\cdots,\tau\}}\sum_{i\in\mathcal{S}_{t^\mathrm{p}}}\tau\ddot{p}_i\beta_{il_k^\mathrm{M}}^\mathrm{NLOS},
      \end{equation}
      which ensures the minimal received power of the pilot contamination for the CE of $k^\mathrm{th}$ UE.
      $\mathcal{S}_{t^\mathrm{p}}=\{i|t_i^\mathrm{p}\in\mathcal{I}_\mathrm{P}\&t_i^\mathrm{p}=t^\mathrm{p}\}$ is the set of UEs assigned to pilot $t^\mathrm{p}$ and $\mathcal{I}_\mathrm{P}$ is the set of pilot index of each UE.
  \item[\textit{iii})] During the S-AP assignment, if the $l^\mathrm{th}$ AP has not been assigned with any UE using pilot $t^\mathrm{P}$, it will serve one of these UEs, e.g., the $i^\mathrm{th}$ UE (with $t_i^\mathrm{P}=t^\mathrm{P}$), which has the best large-scale fading coefficient but only under the condition that $\beta_{il}-\beta_{il_i^\mathrm{M}}\geq\eta$ where $\eta<0$ is a threshold. This assignment method will guarantee that each S-AP serves at most one UE over each orthogonal pilots.
  \item[\textit{iv})] After the completion of AP cluster formation (Steps \textit{i} and \textit{iii}) and pilot assignment (Step \textit{ii}) for each UE, the transmit power of each UE is updated according to the large-scale fading coefficients values obtained from all its serving APs, which will be used to carry out the next round of pilot assignment and AP cluster formation.
\end{enumerate}
\begin{algorithm}[tp]
\caption{Joint AP Cluster Formation, Pilot Assignment, and UL Power Control.}\label{alg1}
\KwIn{$\{\beta_{kl}\}$, $\{\kappa_{kl}\}$, $\mathcal{L}(\bar{d})$, $\{\mathcal{S}_{t_k^\mathrm{p}}=\emptyset\}$}
\KwOut{$\mathcal{I}_\mathrm{P}$, $\{\mathcal{M}_k\}$, $\{\mathcal{M}_k^\mathrm{P}\}$ ,$\{\ddot{p}_k\}$}
\For{$m=1:M$}
{
\For{$k=1:K$}
{
\If{$m==1$}
{$l'=\arg\max_{l\in\mathcal{L}(\bar{d})}\beta_{kl}$;
\\$\mathcal{M}_k^{\mathrm{P}}\leftarrow(l_k^\mathrm{M}=l')$;
\\$\vect{D}_{kl'}\leftarrow\vect{I}_N$;}
\If{$k\leq\tau$}
{$\mathcal{I}_\mathrm{P}\leftarrow\mathcal{I}_\mathrm{P}\cup\{t_k^\mathrm{p}=k\}$;
\\\Else{$\mathcal{I}_\mathrm{P}\leftarrow\mathcal{I}_\mathrm{P}
\cup\{t_k^\mathrm{p}=\arg\min_{t^\mathrm{p}\in\{1,2,\cdots,\tau\}}\sum_{i\in\mathcal{S}_{t^\mathrm{p}}}\tau\ddot{p}_i\beta_{il_k^\mathrm{M}}^\mathrm{NLOS}\}$;}}
}
\For{$l=1:L$}
{
\For{$t^\mathrm{p}=1:\tau$}
{
\If{$\sum_{i\in\mathcal{S}_{t^\mathrm{p}}}\tr(\vect{D}_{il})==0$}
{$i'=\arg\max_{i\in\mathcal{S}_{t^\mathrm{p}}}\ddot{p}_i\beta_{il}$;
\\\If{$\beta_{i'l}-\beta_{i'l_{i'}^\mathrm{M}}\geq\eta$}
{$\vect{D}_{i'l}\leftarrow\vect{I}_N$;}
}
}
}
\For{$k=1:K$}
{$\mathcal{M}_k\leftarrow\{l|\vect{D}_{kl}=\vect{I}_N\And l=1,2,\cdots,L\}$;
\\$\ddot{p}_k\leftarrow{p}(1-\rho_\mathrm{da})\min_{i\in\mathcal{Q}_k}(\sum_{l\in\mathcal{M}_i}\beta_{il})^\nu/(\sum_{l\in\mathcal{M}_k}\beta_{kl})^\nu$;}
\If{$m<M$}
{$\mathcal{I}_\mathrm{P}\leftarrow\emptyset$;
\\$(\mathcal{M}_k^\mathrm{S}=\mathcal{M}_k-\mathcal{M}_k^\mathrm{p})\leftarrow\emptyset$;}
}
{\bf{final}};\\
\end{algorithm}
The detailed pseudo code for the above algorithm is given as Algorithm \ref{alg1}.
From the code given in its line $19$, where the transmit power of each UE is updated based on the large-scale fading coefficients, we get the following conclusions: \textit{i}) The CC of our proposed power control is proportional to $|\mathcal{Q}_k|$ and $|\mathcal{M}_k|$ which is independent of $K$ and $L$. Recalling that the CC in \cite{Chen2020Structured} and \cite{Nikbakht2020} is proportional to $K$ and $L$, respectively, one can obtain that our proposed power control scheme is indeed scalable when $K\rightarrow\infty$ and $L\rightarrow\infty$, which is a big advantage for the CC of UEs' power control; \textit{ii}) If $k=\min_{i\in\mathcal{Q}_k}(\sum_{l\in\mathcal{M}_i}\beta_{il})^\nu$ which means $k^\mathrm{th}$ UE is a cell-edge UE, one can obtain $\ddot{p}_k=p(1-\rho_\mathrm{da})$ and $\ddot{p}_i<p(1-\rho_\mathrm{da})$ for $\forall i\in\mathcal{Q}_k\backslash{k}$, which guarantees QoS fairness for all accessing UEs.
Following the above observations, it is clear that the proposed algorithm achieves simultaneous scalability and QoS fairness for all UEs of the considered SCF-mMIMO and to the best of our knowledge, this is the first time that such contribution is being published in the open technical literature.

\subsection{Complexity Analysis for Algorithm \ref{alg1}}
For the P-AP assignment over $K$ accessing UEs, where each UE independently chooses one AP from the set $\mathcal{L}(\bar{d})$ as its P-AP, the complexity becomes $\mathcal{O}(K|\mathcal{L}(\bar{d})|)$. For the S-AP assignment, where each AP goes through $\tau$ different pilot indexes, the complexity becomes $\mathcal{O}(L\tau)$.

For the pilot assignment over $K$ accessing UEs, where the pilot indexes of first $\tau$ UEs are sequentially assigned with $\{1,2,\cdots,\tau\}$ and the pilot indexes of the last $K-\tau$ UEs are assigned by going through $\tau$ different pilot indexes, the complexity for the pilot assignment is $\mathcal{O}(\tau+(K-\tau)\tau)$.

For the power control over $K$ accessing UEs where $i^\mathrm{th}$ UE's power should be updated according to the large-scale fading of $|\mathcal{Q}_i|$ UEs, the complexity for the power control is $\mathcal{O}\big(\sum_{i=1}^{K}|\mathcal{Q}_i|\big)$.
Thus the total complexity of Algorithm \ref{alg1} becomes $\mathcal{O}\big(K|\mathcal{L}(\bar{d})|+(K-\tau+L+1)\tau+\sum_{i=1}^{K}|\mathcal{Q}_i|\big)$.

\section{Performance Evaluation Results and Discussion}\label{section:simulation}
The proposed SCF-mMIMO system has been implemented in software and its performance has been obtained by means of computer simulation experiments so that the accuracy of the theoretical analysis can be also verified.
Without loss of generality, it is assumed that there are $L=64$ APs which are uniformly distributed within a $1\times1$ $\mathrm{km}^2$ area, and the number of UEs is $K=40$. Since ULA is used at every AP, $[\vect{R}_{kl}]_{n_1n_2}$ can be computed as \cite[Eq. (2.23)]{Emil2017mMnetworks}
\begin{align}
\notag[\vect{R}_{kl}]_{n_1n_2}
&=
\int_{-20\sigma_\varphi}^{20\sigma_\varphi}\exp\left({\mathrm{j}2\pi \frac{d_\mathrm{H}}{\lambda}(n_1-n_2)\sin(\varphi+\delta)}\right)
\\&\quad\times\beta_{kl}^\mathrm{NLOS}\frac{1}{\sqrt{2\pi}\sigma_\varphi}\exp\left({-\frac{\delta^2}{2\sigma_\varphi^2}}\right)\mathrm{d}\delta,
\end{align}
where $d_\mathrm{H}=0.5\lambda$ is the distance between the antennas and $\lambda$ is the carrier wavelength.
Moreover, $\varphi=\theta_{kl}$ is a deterministic nominal angle between the $k^\mathrm{th}$ UE and the $l^\mathrm{th}$ AP \cite{Z.Wang2020uplink}, and $\delta\sim\mathcal{N}(0,\sigma_\varphi^2)$ is a random deviation from $\varphi$. The Rician factor between $k^\mathrm{UE}$ and $l^\mathrm{AP}$ is given as $\kappa_{kl}[\mathrm{dB}]=13-0.03d_{kl}$ \cite{Ozdogan2019Massive},
and the large-scale fading is modeled as \cite[(37)]{Emil2020making}
\begin{equation}
\beta_{kl}[\mathrm{dB}]=-30.5-36.7\log_{10}\left(\frac{d_{kl}}{1\,\mathrm{m}}\right)+F_{kl},
\end{equation}
where $d_{kl}$ and $F_{kl}\sim\mathcal{N}(0,4^2)$ are the distance and shadow fading between $k^\mathrm{th}$ UE and $l^\mathrm{th}$ AP, respectively.
The values of the various operational parameters used in our computer simulation experiments are in accordance with those used in \cite{Emil2020making} and have been summarized in Table \ref{system parameter}.
\begin{table}[h]
\renewcommand{\arraystretch}{1.}
\caption{\textsc{Operational Parameters Values Used for the Simulations}}
\vskip 1mm
\label{system parameter}
\centering
\begin{tabular}{c|c}
\hline
\bfseries Operational Parameter&\bfseries Value
\\\hline
Transmission bandwidth: $B$&$20$ MHz
\\\hline
Channel coherence bandwidth: $B_\mathrm{c}$&$100$ kHz
\\\hline
Channel coherence time: $T_\mathrm{c}$&$2$ ms
\\\hline
Coherence block: $\tau_c$&$200$ symbols
\\\hline
Pilots block: $\tau$&$10$ symbols
\\\hline
Transmit power: $p$&$100$ mW
\\\hline
Noise figure: $\iota$&$5$ dB
\\\hline
Noise power: $\sigma^2=-174+10\log_{10}(B)+\iota$&$-96$ dBm
\\\hline
Angular standard deviation (ASD): $\sigma_\varphi$&$15\degree$
\\\hline
Threshold: $\eta$&$-20$ dB
\\\hline
\end{tabular}\vspace{-0.1cm}
\end{table}

In this section, firstly the SE performance comparison for different Q-bit and channel conditions will be presented in subsection \ref{sub-A}. In subsection \ref{sub-B}, by using Eqs. \eqref{SNR-d} and \eqref{UL-SE-C}, the SE performance by using our proposed scalable detection methods will be compared with the conventional unscalable methods.
In subsection \ref{sub-C}, the SE performance for our proposed pilot assignment and power control algorithm will be compared with the conventional random pilot assignment and equal power control policies.

\subsection{UL-SE Performance Comparison between Distributed and Centralized Schemes}\label{sub-A}
\begin{figure}
\begin{center}
\includegraphics[width=0.9\columnwidth]{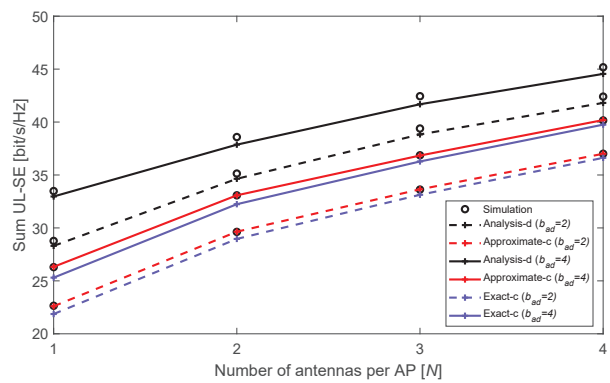} 
\captionsetup{font={normal}}
\caption{Sum SE with different connection schemes and number of antennas per AP.}\label{sum_UL-SE}
\end{center} \vskip-8mm
\end{figure}

In Fig. \ref{sum_UL-SE}, the sum SE performance with distributed and centralized schemes and for different Q-bit is plotted against the number of antennas per AP employing MRC detection.
Specifically, we set $b_\mathrm{da}=1$, $b_\mathrm{ad}=2,\,4$.
It is clear from these performance results that the more Q-bit are used at each AP, the higher the sum SE performance becomes.

The line `Analysis-d' corresponds to the SE performance for the distributed scheme (i.e., \eqref{UL-SE-MRC}), while lines `Exact-c' and `Approximate-c' represent the SE performance for the centralized scheme (i.e., \eqref{UL-SE-C} and \eqref{UL-SE-C-approx-close}).
It is clear that the simulation results match well with their corresponding analytical results, thus confirming the correctness of Theorems \ref{Theorem-2} and \ref{Theorem-3}.
For the centralized scheme, it is noted that the SE performance gap between the accurate line and the approximate line is very small, thus making it reasonable to approximate \eqref{UL-SE-C} with \eqref{UL-SE-C-approx-close}.
\begin{figure*}
\centering
\captionsetup{font={normal}}
\subfigure[]{
\begin{minipage}[t]{0.45\linewidth}
\centering
\includegraphics[width=0.95\linewidth]{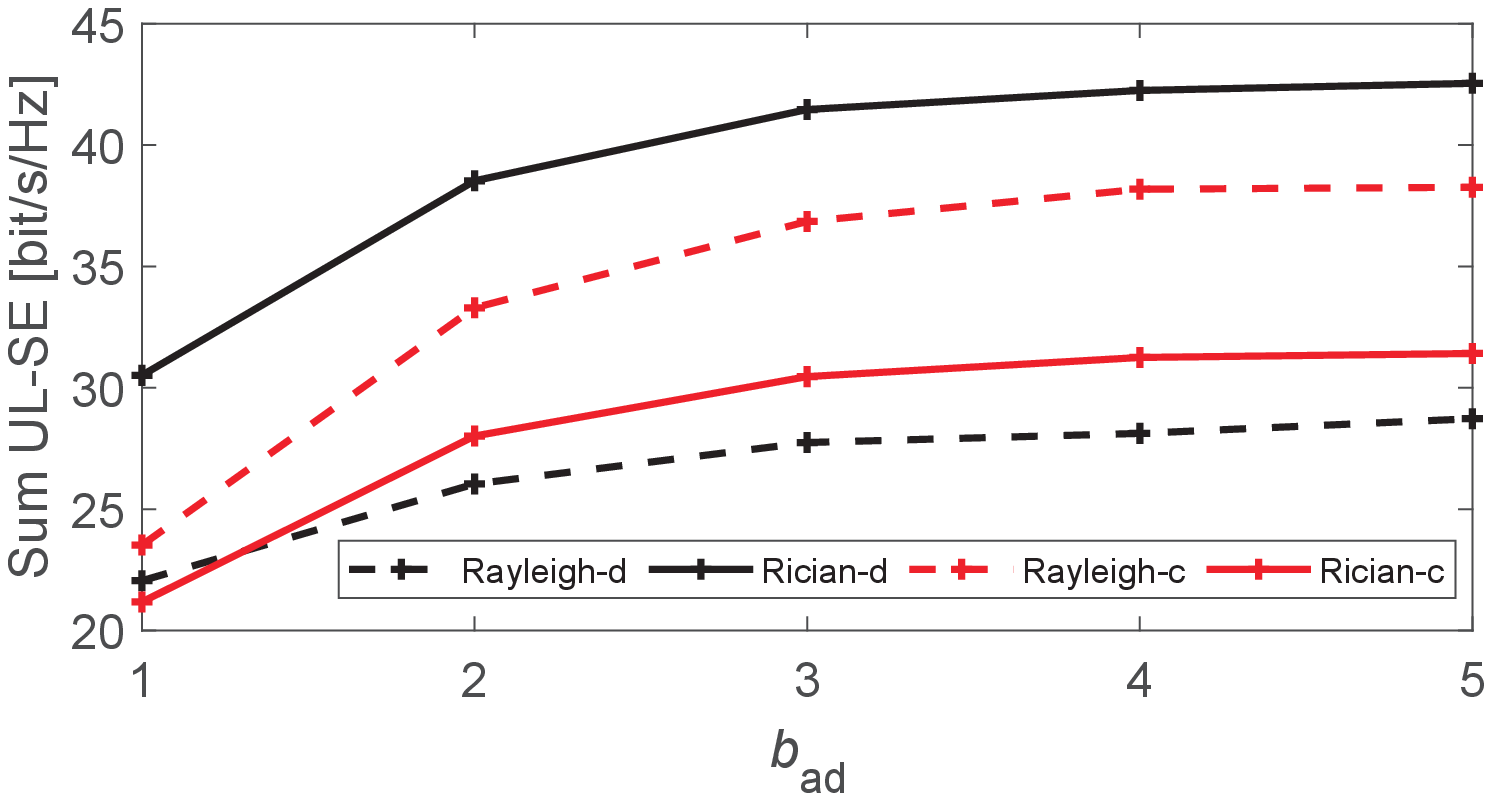}
\end{minipage}
}
\captionsetup{font={normal}}
\subfigure[]{
\begin{minipage}[t]{0.46\linewidth}
\centering
\includegraphics[width=0.95\linewidth]{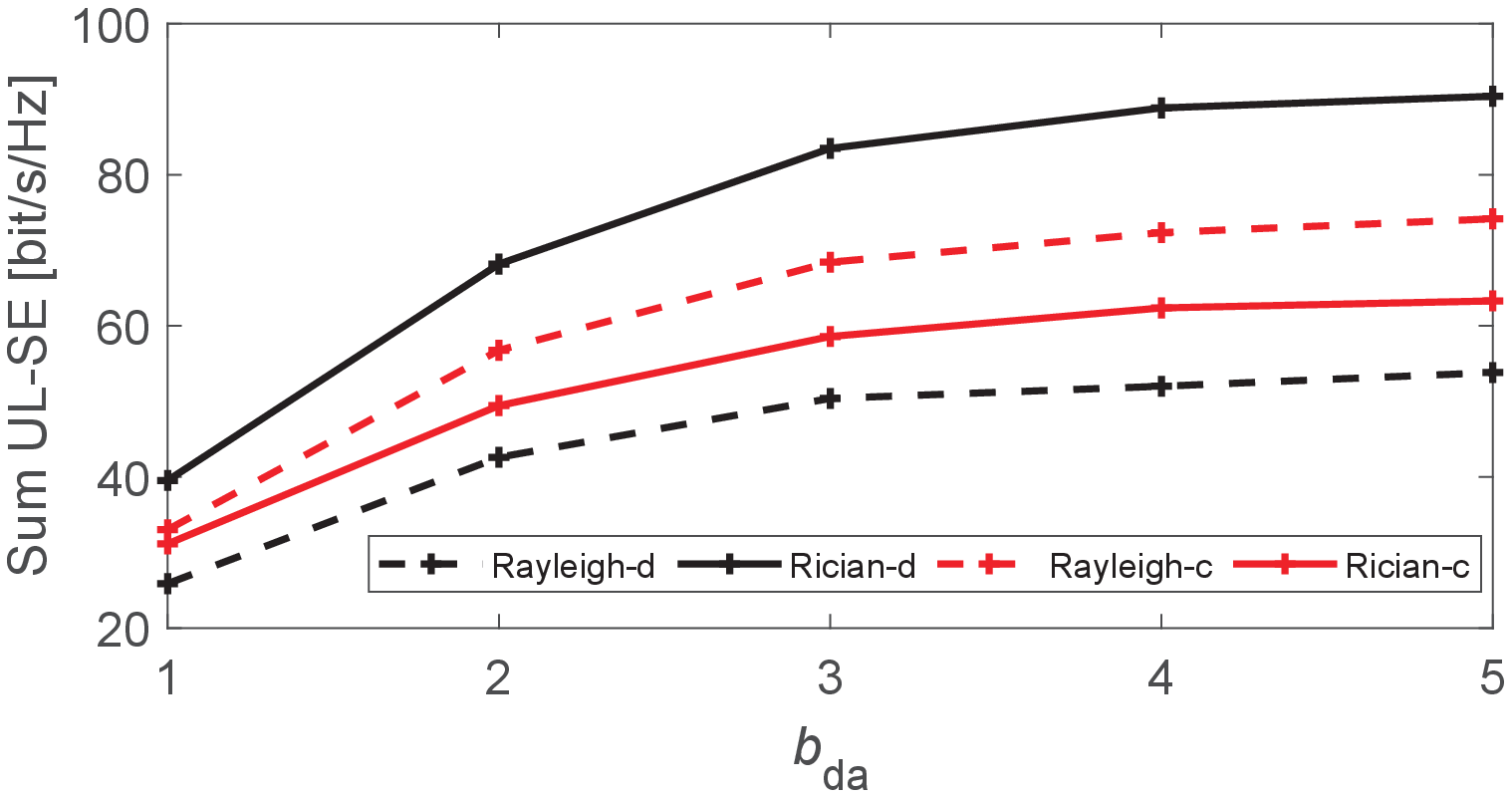}
\end{minipage}
}
\centering
\captionsetup{font={normal}}
\caption{Sum SE with different channel fading and Q-bit of ADCs/DACs. (a) $b_\mathrm{da}=1$. (b) $b_\mathrm{ad}=2$.}
\label{sum_UL-SE-b}
\end{figure*}

In Fig. \ref{sum_UL-SE-b}, the sum SE performance for different connection schemes and channel fading conditions are plotted against the Q-bit where $N=2$ and MRC detector is used.
Specifically, in Fig. \ref{sum_UL-SE-b}(a), we set $b_\mathrm{da}=1$ and increase $b_\mathrm{ad}$ from $1$ to $5$; In Fig. \ref{sum_UL-SE-b}(b), we set $b_\mathrm{ad}=2$ and increase $b_\mathrm{da}$ from $1$ to $5$.
The obtained performance results clearly show that the increase of SE tails off for $b_\mathrm{ad}>3$ or $b_\mathrm{da}>4$.
It is noted that from Fig. \ref{sum_UL-SE-b}, in the presence of Rayleigh fading, our centralized scheme outperforms the distributed scheme which coincide with the result presented in \cite{Emil2020making}. However, in the presence of Rician fading channels, our simulation results show that the distributed scheme outperforms the centralized scheme. This happens because the LOS path of Rician fading is deterministic and assumed to be known at the CPUs, which makes the unknown channel gain $\mathbb{E}[|\vect{a}_k^\mathrm{H}(\vect{g}_{kk}-\mathbb{E}[\vect{g}_{kk}])|^2]$ relatively smaller than the average channel gain $|\vect{a}_k^\mathrm{H}\mathbb{E}[\vect{g}_{kk}]|^2$ in \eqref{eq:UL-d}.
It is noted from the results presented in Fig. \ref{sum_UL-SE-b} that, when $b_\mathrm{ad}$ and $b_\mathrm{da}$ take values greater than $5$, the SE performance under Rayleigh is identical with that reported in \cite[Fig. 3]{Emil2020making}. This observation confirms the correctness of the analysis as well as the accuracy of our performance evaluation results even for the special case of Rayleigh fading presented in \cite{Emil2020making}.

\subsection{UL-SE Performance Comparison between Different Detection Methods}
\label{sub-B}
\begin{figure*}
\centering
\captionsetup{font={normal}}
\subfigure[]{
\begin{minipage}[t]{0.45\linewidth}
\centering
\includegraphics[width=0.95\linewidth]{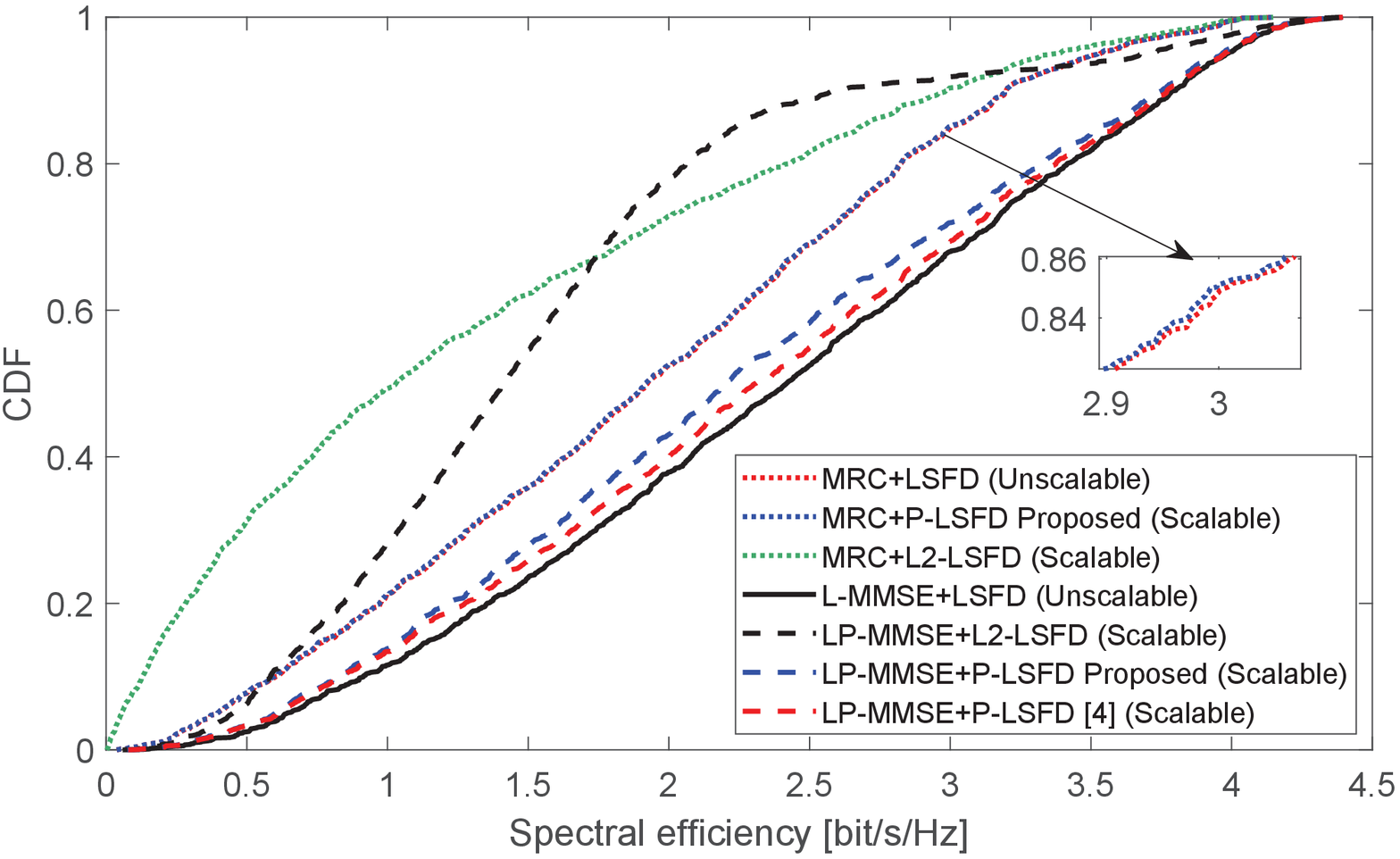}
\end{minipage}
}
\captionsetup{font={normal}}
\subfigure[]{
\begin{minipage}[t]{0.45\linewidth}
\centering
\includegraphics[width=0.95\linewidth]{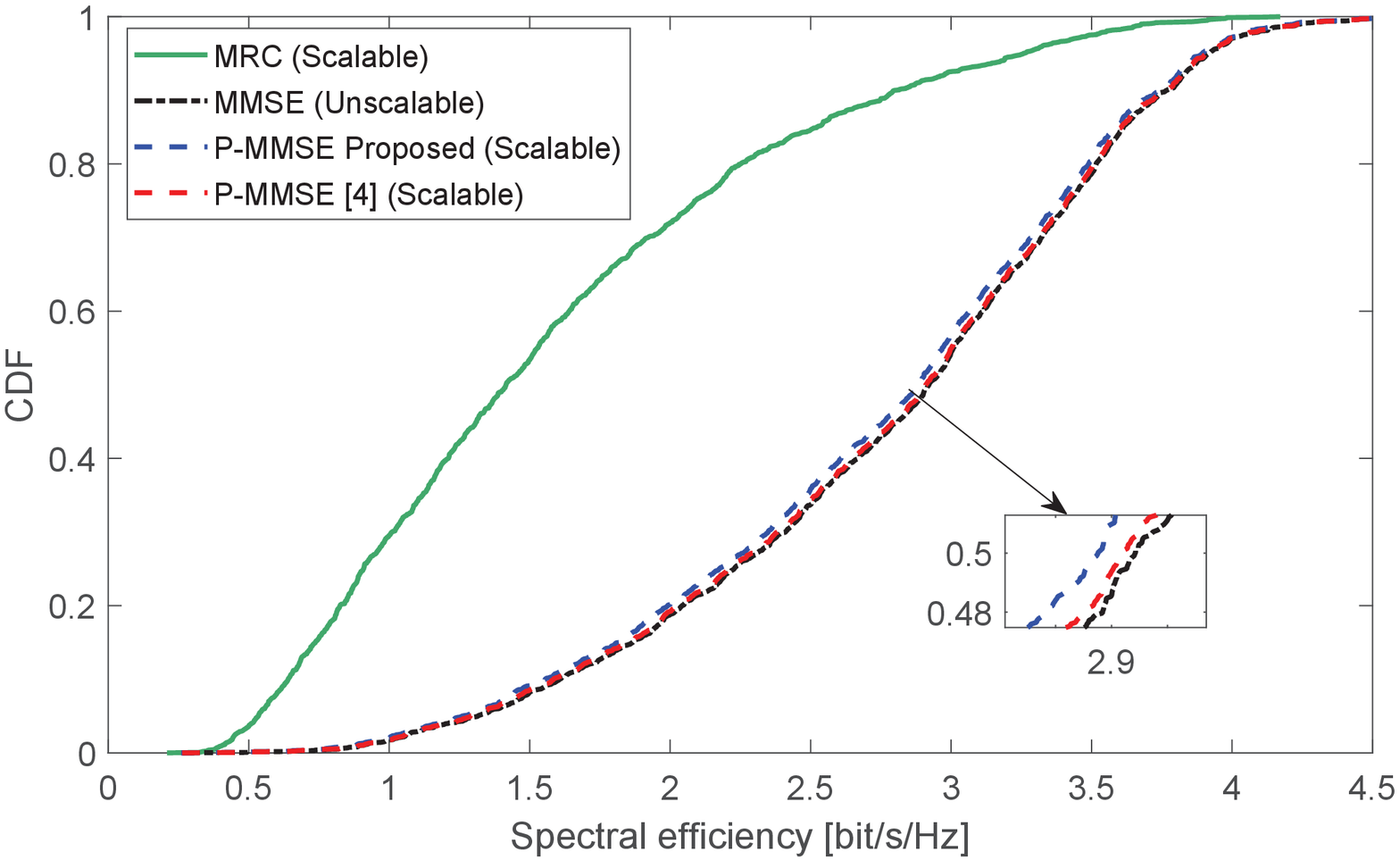}
\end{minipage}
}
\centering
\captionsetup{font={normal}}
\caption{SE per UE with different scalable and unscalable detection methods. (a) Distributed scheme. (b) Centralized scheme.}
\label{UL-SE_CDF}
\end{figure*}

The cumulative distribution functions (CDFs) of SE for distributed and centralized schemes are depicted in Fig. \ref{UL-SE_CDF} where $N=3$ and $K=80$. According to the SINR term in \eqref{SNR-d}, the SE performance for our proposed scalable MRC+P-LSFD detection method and LP-MMSE+P-LSFD detection method are compared two benchmarks:
a) MRC $+$ LSFD \cite[(21)]{Emil2020making} (dotted red line);
b) L-MMSE $+$ LSFD (solid black line).
In contrast to LSFD vector, the CC for calculating our P-LSFD is proportional to $|\mathcal{Q}_k|$ which is not grown arbitrarily large as $K\rightarrow\infty$.
It is shown that the SE performance of our proposed two scalable detection methods are very close to their corresponding unscalable methods.

Another observation from Fig. \ref{UL-SE_CDF}(a) is that, as compared to the original LP-MMSE vector in \cite{Emil2020scalable}, our proposed LP-MMSE detector yields very similar SE performance while requiring significantly lower CC.
Note that the L2-LSFD vector in Fig. \ref{UL-SE_CDF}(a) is set as $\vect{1}_{|\mathcal{M}_k|}$ that is also a scalable LSFD vector
independent of the CSIs \cite{Emil2020scalable}.
It becomes clear that the SE performance is highly influenced by the specific form of the scalable LSFD vector.

According to the SE expression in \eqref{UL-SE-C}, the CDF curve of our proposed P-MMSE detector are compared with original P-MMSE detector in \cite{Emil2020scalable} and the conventional MMSE detector which is illustrated in Fig. \ref{UL-SE_CDF}(b). It is obvious that the SE performance using MMSE detector achieves much higher SE performance than the MRC does. Similarly to the observations from Fig. \ref{UL-SE_CDF}(a), again it can be concluded that the SE performance gap between the three different MMSE detectors is very small while our proposed P-MMSE vector has the smallest CC.

\subsection{UL-SE Performance with Different Pilot Assignment and Power Control Strategies}\label{sub-C}
\begin{figure*}
\centering
\captionsetup{font={normal}}
\subfigure[]{
\begin{minipage}[t]{0.45\linewidth}
\centering
\includegraphics[width=0.95\linewidth]{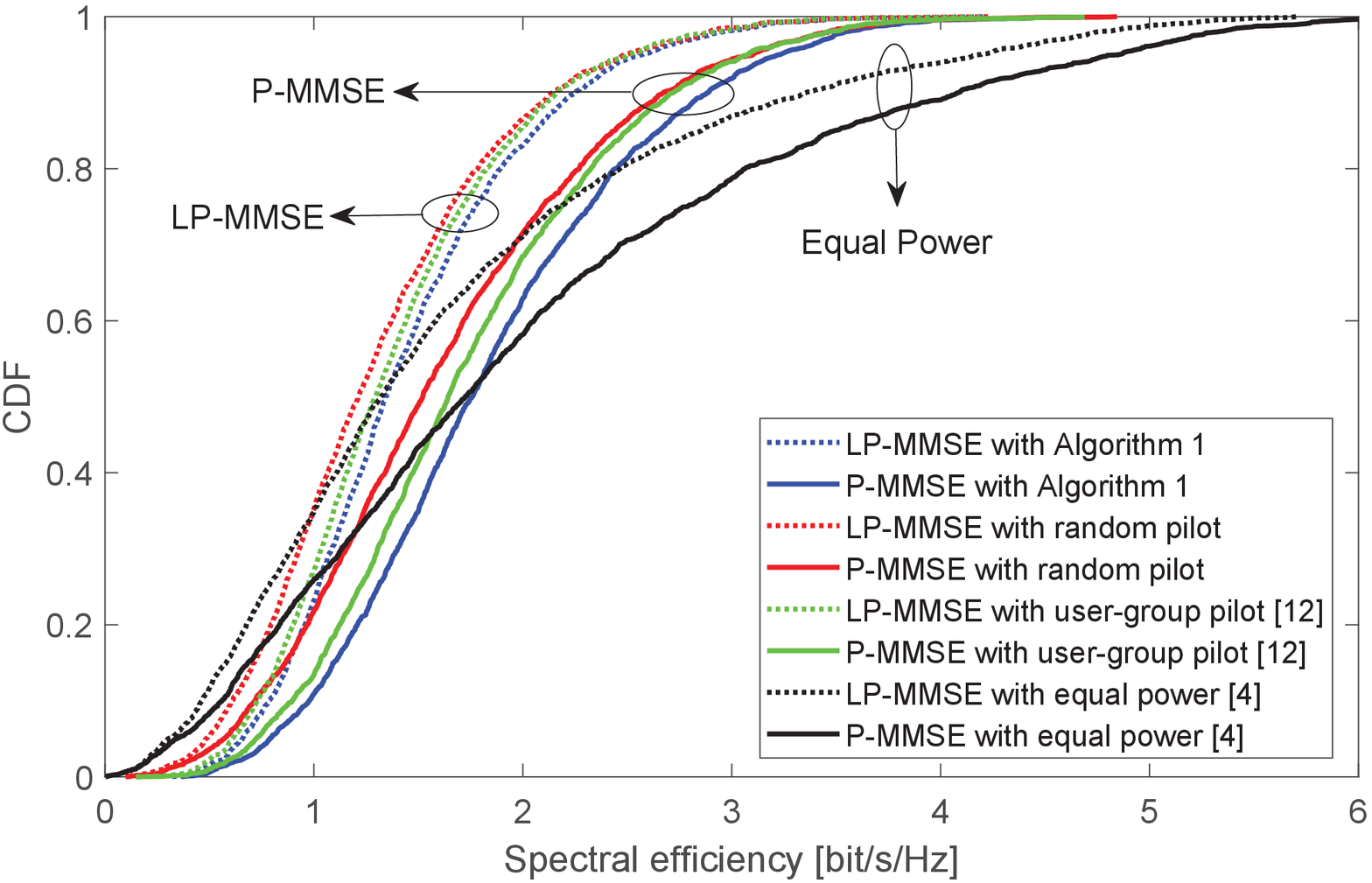}
\end{minipage}
}
\captionsetup{font={normal}}
\subfigure[]{
\begin{minipage}[t]{0.45\linewidth}
\centering
\includegraphics[width=0.95\linewidth]{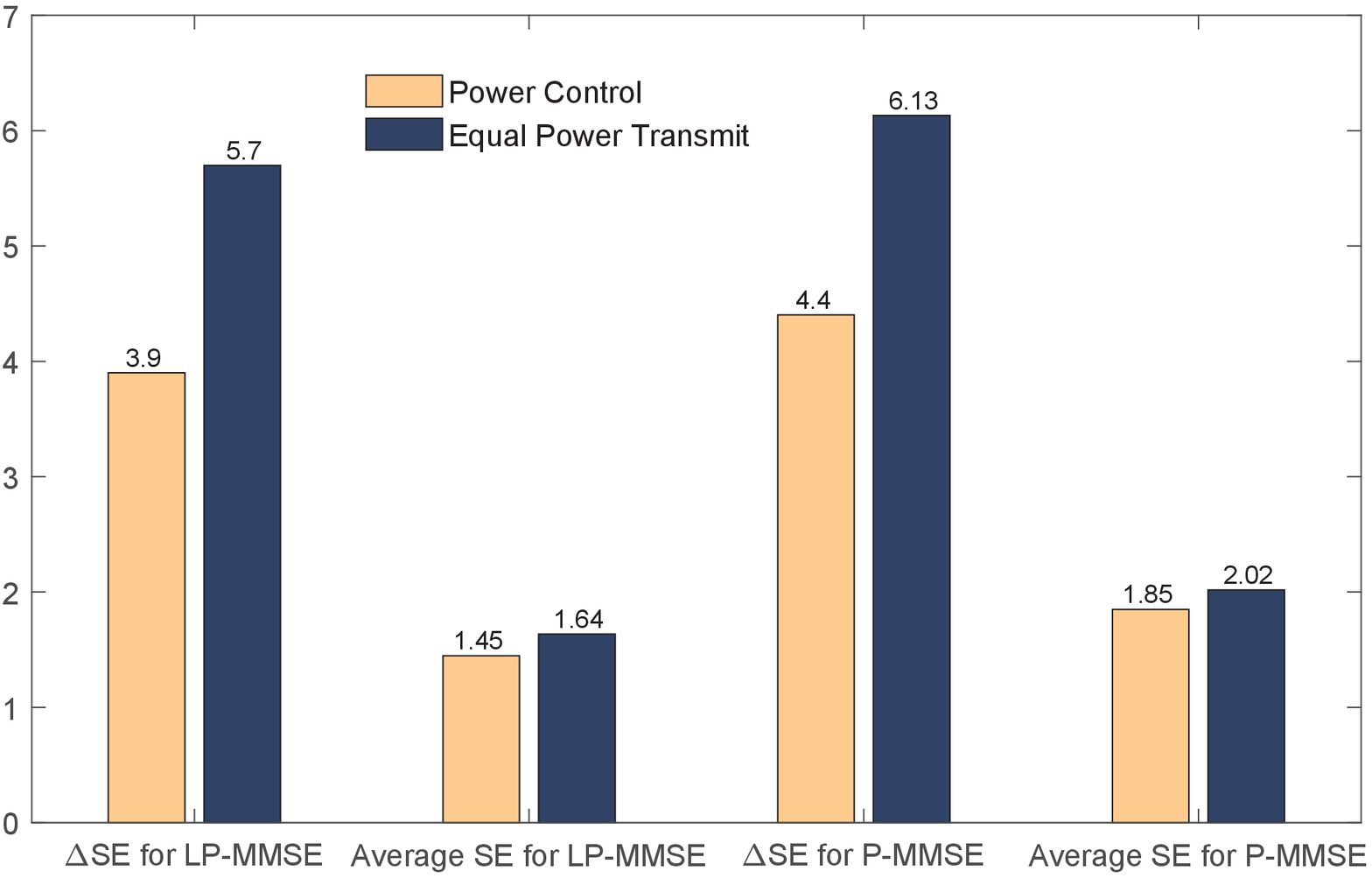}
\end{minipage}
}
\centering
\captionsetup{font={normal}}
\caption{SE performance evaluation of Algorithm \ref{alg1}: (a) SE performance comparison between different pilot assignment and UL power control strategies. (b) Tradeoff between QoS fairness and average SE.}
\label{UL-SE_CDF_alg}
\vspace{-0.2 in}
\end{figure*}
The SE performance for Algorithm \ref{alg1} is depicted in Fig. \ref{UL-SE_CDF_alg} where $N=3$ and $b_\mathrm{ad}=b_\mathrm{da}=4$. Since the LOS path is deterministic and the CE accuracy is highly influenced by the number of UEs sharing the same pilot matrix, we specifically choose correlated Rayleigh fading channel and set $K=150$. In this way, a more clear SE performance comparison can be made between different pilot assignment strategies.

In Fig. \ref{UL-SE_CDF_alg}(a), the CDF curves of the SE performance using Algorithm \ref{alg1} with $\nu=0.8$ are plotted and compared with the equivalent performance of different pilot assignment strategies (i.e., random pilot assignment and user-group based pilot assignment in \cite{Chen2020Structured}) and UL equal power transmit strategy in \cite{Emil2020scalable}, where two scalable LP-MMSE and P-MMSE detection methods are adopted in distributed and centralized schemes, respectively.
It is clear from these results that our proposed pilot assignment scheme yields much higher SE performance enhancement as compared to both the random pilot assignment and user-group based pilot assignment while requiring significantly lower CC than the user-group based pilot assignment.

It can also be observed from both Figs. \ref{UL-SE_CDF_alg}(a) and \ref{UL-SE_CDF_alg}(b) that our proposed power control scheme achieves better QoS fairness as compared with the equal power transmit strategy in \cite{Emil2020scalable}.
Specifically in Figs. \ref{UL-SE_CDF_alg}(b) where the QoS fairness (measured by $\Delta\mathrm{SE}=\mathrm{SE_{max}-\mathrm{SE_{min}}}$) and average SE performance for different detection schemes are presented, it is clear that our proposed scalable power control strategy achieves $31.6\%$ and $28.2\%$ fairness improvement at the expense of $11.6\%$ and $8.4\%$ loss of average SE for distributed and centralized schemes, respectively.
\begin{figure}
\begin{center}
\includegraphics[width=0.9\columnwidth]{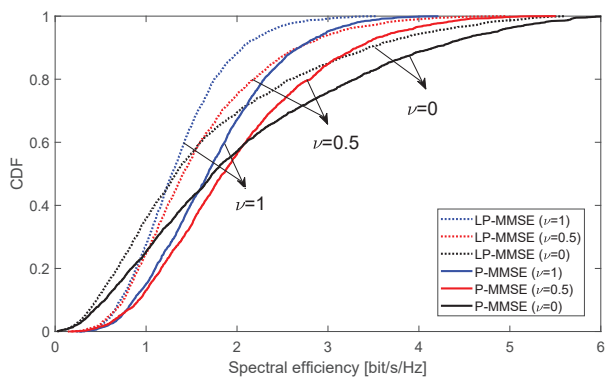} 
\captionsetup{font={normal}}
\caption{SE Performance variations for different values of $\nu$.}\label{UL-SE_CDF_nu}
\end{center} \vskip-8mm
\end{figure}
A similar trend can also be observed from Fig. \ref{UL-SE_CDF_nu} where the tradeoff between the QoS fairness and average SE are illustrated with the increase of $\nu$ from $0$ (equal power transmit strategy) to $1$. It is noted that, with the increase of $\nu$, a set of progressively compressed SE CDF curves are presented, which means that any desired balance we want between the QoS fairness and the average SE can be obtained by simply adjusting the value of $\nu$.

\section{Conclusion}\label{section:conclusion}
In this paper, we have presented an analytical framework for a TDD SCF-mMIMO system employing finite resolution ADCs/DACs and operating in correlated Rician fading, and have made several novel contributions.
Firstly, the SE expressions for both distributed and centralized scheme have been derived by using MRC detection, and was shown that SE performance growth tails off for $b_\mathrm{ad}>3$ or $b_\mathrm{da}>4$.
Secondly, in order to achieve the scalability of signal detection, the LP-MMSE and P-MMSE detectors and P-LSFD weighting vector are proposed whose CCs are greatly reduced and achieve similar SE performance as the unscalable methods, namely L-MMSE, MMSE, and LSFD.
Thirdly, we have introduced a scalable algorithm for the considered system to carry out the AP cluster formation, pilot assignment, and power control for each UE.
Various analytical performance evaluation results complemented by equivalent computer simulated results have shown that the proposed algorithm can greatly mitigate the pilot contamination while achieving the goal of QoS fairness for all accessing UEs.

\begin{appendices}
\section{Proof of Theorem \ref{Theorem-1}}\label{Appendix:Theorem-1}
\subsubsection{Derivation of \eqref{expectation1-a-1}}
For $\forall{i}\in\mathcal{P}_k$ and $l_1=l_2$, one can obtain \cite{Ozdogan2019Massive}
\begin{align}\label{expectation1-a}
\mathbb{E}[\vect{\hat{h}}_{kl_1}^\mathrm{H}\vect{h}_{il_1}\vect{h}_{il_2}^\mathrm{H}\vect{\hat{h}}_{kl_2}]
=\mathbb{E}[|\vect{\hat{h}}_{kl_1}^\mathrm{H}\vect{\hat{h}}_{il_1}|^2]
+\mathbb{E}[|\vect{\hat{h}}_{kl_1}^\mathrm{H}\vect{\tilde{h}}_{il_1}|^2].
\end{align}
Since $\vect{\Psi}_{t_k^\mathrm{p},l}=\vect{\Psi}_{t_i^\mathrm{p},l}$ and $\vect{z}_{t_k^\mathrm{p},l}^\mathrm{w}=\vect{z}_{t_i^\mathrm{p},l}^\mathrm{w}$ for $\forall{i}\in\mathcal{P}_k$, $\vect{\hat{h}}_{kl_1}$ and $\vect{\hat{h}}_{il_1}$ can be reformulated as
\begin{align}
\notag\vect{\hat{h}}_{kl_1}&=\vect{\bar{h}}_{kl_1}+(1-\rho_\mathrm{ad})\sqrt{\ddot{p}_k\tau}\vect{R}_{kl_1}\vect{\Psi}_{t_k^\mathrm{p},l_1}^{-1}\vect{\Psi}_{t_k^\mathrm{p},l_1}^\frac{1}{2}\vect{\Psi}_{t_k^\mathrm{p},l_1}^{-\frac{1}{2}}\vect{z}_{t_k^\mathrm{p},l_1}^\mathrm{w}
\\&=\vect{\bar{h}}_{kl_1}+(1-\rho_\mathrm{ad})\sqrt{\ddot{p}_k\tau}\vect{R}_{kl_1}\vect{\Psi}_{t_k^\mathrm{p},l_1}^{-\frac{1}{2}}\vect{w}_{l_1},
\\\vect{\hat{h}}_{il_1}&=\vect{\bar{h}}_{il_1}+(1-\rho_\mathrm{ad})\sqrt{\ddot{p}_i\tau}\vect{R}_{il_1}\vect{\Psi}_{t_k^\mathrm{p},l_1}^{-\frac{1}{2}}\vect{w}_{l_1},
\end{align}
where $\vect{w}_{l_1}\sim\mathcal{CN}(\vect{0},\vect{I}_N)$.
By using \cite[Lemma 5]{Ozdogan2019Massive}, $\mathbb{E}[|\vect{\hat{h}}_{kl_1}^\mathrm{H}\vect{\hat{h}}_{il_1}|^2]$ in \eqref{expectation1-a} can be calculated as \eqref{ex-correlation1}.
\begin{figure*}[!t]
\begin{align}\label{ex-correlation1}
\notag\mathbb{E}[|\vect{\hat{h}}_{kl_1}^\mathrm{H}\vect{\hat{h}}_{il_1}|^2]&=|\vect{\bar{h}}_{kl_1}^\mathrm{H}\vect{\bar{h}}_{il_1}|^2
+(1-\rho_\mathrm{ad})^4\tau^2\ddot{p}_k\ddot{p}_i|\tr(\vect{R}_{il_1}\vect{\Psi}_{t_k^\mathrm{p},l_1}^{-1}\vect{R}_{kl_1})|^2
+(1-\rho_\mathrm{ad})^4\tau^2\ddot{p}_k\ddot{p}_i\tr(\vect{R}_{il_1}\vect{\Psi}_{t_k^\mathrm{p},l_1}^{-1}\vect{R}_{il_1}\vect{R}_{kl_1}\vect{\Psi}_{t_k^\mathrm{p},l_1}^{-1}\vect{R}_{kl_1})
\\\notag&\quad+(1-\rho_\mathrm{ad})^2\tau\ddot{p}_k\vect{\bar{h}}_{il_1}^\mathrm{H}\vect{R}_{kl_1}\vect{\Psi}_{t_k^\mathrm{p},l_1}^{-1}\vect{R}_{kl_1}\vect{\bar{h}}_{il_1}
+(1-\rho_\mathrm{ad})^2\tau\ddot{p}_i\vect{\bar{h}}_{kl_1}^\mathrm{H}\vect{R}_{il_1}\vect{\Psi}_{t_k^\mathrm{p},l_1}^{-1}\vect{R}_{il_1}\vect{\bar{h}}_{kl_1}
\\&\quad+2(1-\rho_\mathrm{ad})^2\tau\sqrt{\ddot{p}_i\ddot{p}_k}\Re\big\{\tr\big(\vect{R}_{il_1}\vect{\Psi}_{t_k^\mathrm{p},l_1}^{-1}\vect{R}_{kl_1}\big)\vect{\bar{h}}_{il_1}^\mathrm{H}\vect{\bar{h}}_{kl_1}\big\}
\end{align}
\hrulefill
\end{figure*}
Since
\begin{equation}
\mathbb{E}\big[\vect{\hat{h}}_{kl_1}\vect{\hat{h}}_{kl_1}^\mathrm{H}\big]=\vect{\bar{h}}_{kl_1}\vect{\bar{h}}_{kl_1}^\mathrm{H}+(1-\rho_\mathrm{ad})^2\tau\ddot{p}_k\vect{R}_{kl_1}\vect{\Psi}_{t_k^\mathrm{p},l_1}^{-1}\vect{R}_{kl_1},
\end{equation}
the second term in \eqref{expectation1-a}, $\mathbb{E}[|\vect{\hat{h}}_{kl_1}^\mathrm{H}\vect{\tilde{h}}_{il_1}|^2]$, can be calculated as
\begin{align}\label{ex-correlation2}
&\notag\mathbb{E}[\vect{\hat{h}}_{kl_1}^\mathrm{H}\vect{\tilde{h}}_{il_1}\vect{\tilde{h}}_{il_1}^\mathrm{H}\vect{\hat{h}}_{kl_1}]=\tr\big(\mathbb{E}\big[\vect{\tilde{h}}_{il_1}\vect{\tilde{h}}_{il_1}^\mathrm{H}\big]\mathbb{E}\big[\vect{\hat{h}}_{kl_1}\vect{\hat{h}}_{kl_1}^\mathrm{H}\big]\big)
\\\notag&=(1-\rho_\mathrm{ad})^2\tau\ddot{p}_k\tr\big(\vect{R}_{il_1}\vect{R}_{kl_1}\vect{\Psi}_{t_k^\mathrm{p},l_1}^{-1}\vect{R}_{kl_1}\big)
\\\notag&\quad-(1-\rho_\mathrm{ad})^4\tau^2\ddot{p}_k\ddot{p}_i\tr(\vect{R}_{il_1}\vect{\Psi}_{t_k^\mathrm{p},l_1}^{-1}\vect{R}_{il_1}\vect{R}_{kl_1}\vect{\Psi}_{t_k^\mathrm{p},l_1}^{-1}\vect{R}_{kl_1})
\\&\quad+\vect{\bar{h}}_{kl_1}^\mathrm{H}\big[\vect{R}_{il_1}-(1-\rho_\mathrm{ad})^2\ddot{p}_i\tau\vect{R}_{il_1}\vect{\Psi}_{t_k^\mathrm{p},l_1}^{-1}\vect{R}_{il_1}\big]\vect{\bar{h}}_{kl_1}.
\end{align}
By substituting \eqref{ex-correlation1} and \eqref{ex-correlation2} into \eqref{expectation1-a}, $\mathbb{E}[\vect{\hat{h}}_{kl_1}^\mathrm{H}\vect{h}_{il_1}\vect{h}_{il_2}^\mathrm{H}\vect{\hat{h}}_{kl_2}]$ can be obtained as in \eqref{expectation1-a-1}.
\subsubsection{Derivation of \eqref{expectation1-a-2}}
For $\forall{i}\in\mathcal{P}_k$ and $l_1\neq l_2$, one can obtain that
\begin{align}\label{expectation1-b}
\notag&\mathbb{E}[\vect{\hat{h}}_{kl_1}^\mathrm{H}\vect{h}_{il_1}\vect{h}_{il_2}^\mathrm{H}\vect{\hat{h}}_{kl_2}]
\\\notag&=\vect{\bar{h}}_{kl_1}^\mathrm{H}\vect{\bar{h}}_{il_1}\vect{\bar{h}}_{il_2}^\mathrm{H}\vect{\bar{h}}_{kl_2}
+\mathbb{E}[\vect{\hat{h}}_{\mathrm{w},kl_1}^\mathrm{H}\vect{h}_{\mathrm{w},il_1}]\mathbb{E}[\vect{h}_{\mathrm{w},il_2}^\mathrm{H}\vect{\hat{h}}_{\mathrm{w},kl_2}]
\\&+\mathbb{E}[\vect{\hat{h}}_{\mathrm{w},kl_1}^\mathrm{H}\vect{h}_{\mathrm{w},il_1}]\vect{\bar{h}}_{il_2}^\mathrm{H}\vect{\bar{h}}_{kl_2}
+\vect{\bar{h}}_{kl_1}^\mathrm{H}\vect{\bar{h}}_{il_1}\mathbb{E}[\vect{h}_{\mathrm{w},il_2}^\mathrm{H}\vect{\hat{h}}_{\mathrm{w},kl_2}].
\end{align}
Since $\vect{\Psi}_{t_k^\mathrm{p},l}=\vect{\Psi}_{t_i^\mathrm{p},l}$ and $\vect{z}_{t_k^\mathrm{p},l}=\vect{z}_{t_i^\mathrm{p},l}$ for $\forall{i}\in\mathcal{P}_k$, the following can be further obtained
\begin{align}\label{ex-correlation3}
\notag&\mathbb{E}[\vect{\hat{h}}_{\mathrm{w},kl_1}^\mathrm{H}\vect{h}_{\mathrm{w},il_1}]=\mathbb{E}[\vect{\hat{h}}_{\mathrm{w},kl_1}^\mathrm{H}\vect{\hat{h}}_{\mathrm{w},il_1}]
=(1-\rho_\mathrm{ad})^2\tau\sqrt{\ddot{p}_k\ddot{p}_i}
\\\notag&\quad\times\tr\big(\mathbb{E}[\vect{R}_{il_1}\vect{\Psi}_{t_k^\mathrm{p},l_1}^{-1}\vect{z}_{t_k^\mathrm{p},l_1}^\mathrm{w}(\vect{z}_{t_k^\mathrm{p},l_1}^\mathrm{w})^\mathrm{H}\vect{\Psi}_{t_k^\mathrm{p},l_1}^{-1}\vect{R}_{kl_1}]\big)
\\&=(1-\rho_\mathrm{ad})^2\tau\sqrt{\ddot{p}_k\ddot{p}_i}\tr\big(\vect{R}_{il_1}\vect{\Psi}_{t_k^\mathrm{p},l_1}^{-1}\vect{R}_{kl_1}\big).
\end{align}
Likewise, we have
\begin{equation}\label{ex-correlation4}
\mathbb{E}[\vect{h}_{\mathrm{w},il_2}^\mathrm{H}\vect{\hat{h}}_{\mathrm{w},kl_2}]=(1-\rho_\mathrm{ad})^2\tau\sqrt{\ddot{p}_k\ddot{p}_i}\tr\big(\vect{R}_{kl_2}\vect{\Psi}_{t_k^\mathrm{p},l_2}^{-1}\vect{R}_{il_2}\big).
\end{equation}
By substituting \eqref{ex-correlation3} and \eqref{ex-correlation4} into \eqref{expectation1-b}, \eqref{expectation1-a-2} can be easily obtained.
\subsubsection{Derivation of \eqref{expectation1-a-3}}
For $\forall{i}\notin\mathcal{P}_k$ and $l_1=l_2$,
\begin{align}\label{expectation1-c}
\notag&\mathbb{E}[\vect{\hat{h}}_{kl_1}^\mathrm{H}\vect{h}_{il_1}\vect{h}_{il_2}^\mathrm{H}\vect{\hat{h}}_{kl_2}]=\mathbb{E}[\vect{\hat{h}}_{kl_1}^\mathrm{H}\vect{h}_{il_1}\vect{h}_{il_1}^\mathrm{H}\vect{\hat{h}}_{kl_1}]
\\\notag&=|\vect{\bar{h}}_{kl_1}^\mathrm{H}\vect{\bar{h}}_{il_1}|^2
+\mathbb{E}[\vect{\hat{h}}_{\mathrm{w},kl_1}^\mathrm{H}\vect{h}_{\mathrm{w},il_1}\vect{h}_{\mathrm{w},il_1}^\mathrm{H}\vect{\hat{h}}_{\mathrm{w},kl_1}]
\\&+\vect{\bar{h}}_{il_1}^\mathrm{H}\mathbb{E}[\vect{\hat{h}}_{\mathrm{w},kl_1}\vect{\hat{h}}_{\mathrm{w},kl_1}^\mathrm{H}]\vect{\bar{h}}_{il_1}
+\vect{\bar{h}}_{kl_1}^\mathrm{H}\mathbb{E}[\vect{h}_{\mathrm{w},il_1}\vect{h}_{\mathrm{w},il_1}^\mathrm{H}]\vect{\bar{h}}_{kl_1},
\end{align}
where
\begin{align}
\notag&\mathbb{E}[\vect{\hat{h}}_{\mathrm{w},kl_1}^\mathrm{H}\vect{h}_{\mathrm{w},il_1}\vect{h}_{\mathrm{w},il_1}^\mathrm{H}\vect{\hat{h}}_{\mathrm{w},kl_1}]
=\tr(\mathbb{E}[\vect{h}_{\mathrm{w},il_1}\vect{h}_{\mathrm{w},il_1}^\mathrm{H}\vect{\hat{h}}_{\mathrm{w},kl_1}\vect{\hat{h}}_{\mathrm{w},kl_1}^\mathrm{H}])
\\&=\tr(\mathbb{E}[\vect{h}_{\mathrm{w},il_1}\vect{h}_{\mathrm{w},il_1}^\mathrm{H}]\mathbb{E}[\vect{\hat{h}}_{\mathrm{w},kl_1}\vect{\hat{h}}_{\mathrm{w},kl_1}^\mathrm{H}]).
\end{align}
\eqref{expectation1-a-3} can be get by substituting $\mathbb{E}[\vect{h}_{\mathrm{w},il_1}\vect{h}_{\mathrm{w},il_1}^\mathrm{H}]=\vect{R}_{il_1}$ and $\mathbb{E}[\vect{\hat{h}}_{\mathrm{w},kl_1}\vect{\hat{h}}_{\mathrm{w},kl_1}^\mathrm{H}]=(1-\rho_\mathrm{ad})^2\tau\ddot{p}_k\vect{R}_{kl_1}\vect{\Psi}_{t_k^\mathrm{p},l_1}^{-1}\vect{R}_{kl_1}$ into \eqref{expectation1-c}.
\subsubsection{Derivation of \eqref{expectation1-a-4}}
For $\forall{i}\notin\mathcal{P}_k$ and $l_1\neq l_2$,
according to \eqref{expectation1-b}, it is clear that
$\mathbb{E}[\vect{\hat{h}}_{kl_1}^\mathrm{H}\vect{h}_{il_1}\vect{h}_{il_2}^\mathrm{H}\vect{\hat{h}}_{kl_2}]=\vect{\bar{h}}_{kl_1}^\mathrm{H}\vect{\bar{h}}_{il_1}\vect{\bar{h}}_{il_2}^\mathrm{H}\vect{\bar{h}}_{kl_2}$,
which completes the proof as it yields \eqref{expectation1-a-4}.

\section{Proof of Theorem \ref{Theorem-2}}\label{Appendix:Theorem-2}
We rewrite the SINR term in \eqref{UL-SE-max} as
\begin{align}
\notag&\mathrm{SINR}_{k,\max}^\mathrm{d}=(1-\rho_\mathrm{ad})^2\ddot{p}_k\mathbb{E}[\vect{g}_{kk}^\mathrm{H}]
\bigg[(1-\rho_\mathrm{ad})^2\sum_{i=1}^{K}\ddot{p}_i\mathbb{E}[\vect{g}_{ki}\vect{g}_{ki}^\mathrm{H}]
\\&+\mathbb{E}[\vect{f}_k\vect{f}_k^\mathrm{H}]
-(1-\rho_\mathrm{ad})^2\ddot{p}_k\mathbb{E}[\vect{g}_{kk}]\mathbb{E}[\vect{g}_{kk}^\mathrm{H}]\bigg]^{-1}\mathbb{E}[\vect{g}_{kk}]
\end{align}

\subsubsection{$\sum_{i=1}^{K}\ddot{p}_i\mathbb{E}[\vect{g}_{ki}\vect{g}_{ki}^\mathrm{H}]$ calculation}
Since with MRC detection $\vect{v}_{kl}=\vect{\hat{h}}_{kl}$, clearly
\begin{equation}
\sum_{i=1}^{K}\ddot{p}_i\mathbb{E}[\vect{g}_{ki}\vect{g}_{ki}^\mathrm{H}]=\sum_{i\in\mathcal{P}_k}\ddot{p}_i\mathbb{E}[\vect{g}_{ki}\vect{g}_{ki}^\mathrm{H}]
+\sum_{i\notin\mathcal{P}_k}\ddot{p}_i\mathbb{E}[\vect{g}_{ki}\vect{g}_{ki}^\mathrm{H}]
\end{equation}
with $\mathbb{E}[\vect{g}_{ki}\vect{g}_{ki}^\mathrm{H}]=\big[\mathbb{E}[\vect{\hat{h}}_{kl_1}^\mathrm{H}\vect{h}_{il_1}\vect{h}_{il_2}^\mathrm{H}\vect{\hat{h}}_{kl_2}]\big]_{|\mathcal{M}_k|}^\mathrm{S}$. According to Theorem \ref{Theorem-1}, one can obtain that
\begin{align}
\notag&\sum_{i\in\mathcal{P}_k}\ddot{p}_i\mathbb{E}[\vect{g}_{ki}\vect{g}_{ki}^\mathrm{H}]=
\sum_{i\in\mathcal{P}_k}\ddot{p}_i\Big[\bm{\lambda}_{k}^i(\bm{\lambda}_{k}^i)^\mathrm{H}
+\vect{b}_k^i(\vect{b}_k^i)^\mathrm{H}
\\&\qquad\qquad\qquad+\vect{b}_k^i(\bm{\lambda}_{k}^i)^\mathrm{H}+\bm{\lambda}_{k}^i(\vect{b}_k^i)^\mathrm{H}
+\big[c_{kl}^i\big]_{|\mathcal{M}_k|}^\Lambda\Big],
\\&\sum_{i\notin\mathcal{P}_k}\ddot{p}_i\mathbb{E}[\vect{g}_{ki}\vect{g}_{ki}^\mathrm{H}]=
\sum_{i\notin\mathcal{P}_k}\ddot{p}_i\Big[\bm{\lambda}_{k}^i(\bm{\lambda}_{k}^i)^\mathrm{H}
+\big[c_{kl}^i\big]_{|\mathcal{M}_k|}^\Lambda\Big].
\end{align}
Thus one can obtain that
\begin{align}\label{ex-Th1-1}
\notag\sum_{i=1}^{K}\ddot{p}_i\mathbb{E}&[\vect{g}_{ki}\vect{g}_{ki}^\mathrm{H}]=
\sum_{i=1}^{K}\ddot{p}_i\Big[\bm{\lambda}_{k}^i(\bm{\lambda}_{k}^i)^\mathrm{H}
+\big[c_{kl}^i\big]_{|\mathcal{M}_k|}^\Lambda\Big]
\\&+\sum_{i\in\mathcal{P}_k}\ddot{p}_i\Big[\vect{b}_k^i(\vect{b}_k^i)^\mathrm{H}
+\vect{b}_k^i(\bm{\lambda}_{k}^i)^\mathrm{H}+\bm{\lambda}_{k}^i(\vect{b}_k^i)^\mathrm{H}\Big].
\end{align}
\subsubsection{$\mathbb{E}[\vect{f}_k\vect{f}_k^\mathrm{H}]$ calculation}
$\vect{n}_l$ in \eqref{n_l-a} can be expressed as
$\vect{n}_l=\vect{n}_{\mathrm{x},l}+\vect{n}_{\mathrm{y},l}$, where $\vect{n}_{\mathrm{x},l}=(1-\rho_\mathrm{ad})\vect{n}_{\mathrm{an},l}+\vect{n}_{\mathrm{ad},l}$ and $\vect{n}_{\mathrm{y},l}=(1-\rho_\mathrm{ad})\sum_{i=1}^{K}\vect{h}_{il}n_{\mathrm{da},i}$ are mutually independent. Since only the term $\vect{n}_{\mathrm{y},l}$ is related to $\vect{h}_{il}$, $\mathbb{E}[\vect{f}_k\vect{f}_k^\mathrm{H}]$ becomes
\begin{equation}\label{Ef_kf_k}
\mathbb{E}[\vect{f}_k\vect{f}_k^\mathrm{H}]=\big[\mathbb{E}[\vect{\hat{h}}_{kl}^\mathrm{H}\vect{n}_{\mathrm{x},l}\vect{n}_{\mathrm{x},l}^\mathrm{H}\vect{\hat{h}}_{kl}]\big]_{|\mathcal{M}_k|}^\Lambda
+\big[\mathbb{E}[\vect{\hat{h}}_{kl_1}^\mathrm{H}\vect{n}_{\mathrm{y},l_1}\vect{n}_{\mathrm{y},l_2}^\mathrm{H}\vect{\hat{h}}_{kl_2}]\big]_{|\mathcal{M}_k|}^\mathrm{S}.
\end{equation}
Since $\mathbb{E}\big[\vect{\hat{h}}_{kl}\vect{\hat{h}}_{kl}^\mathrm{H}\big]=\vect{\bar{h}}_{kl}\vect{\bar{h}}_{kl}^\mathrm{H}+(1-\rho_\mathrm{ad})^2\tau\ddot{p}_k\vect{R}_{kl}\vect{\Psi}_{t_k^\mathrm{p},l}^{-1}\vect{R}_{kl},$
and
\begin{align}
\notag\mathbb{E}\big[\vect{n}_{\mathrm{x},l}&\vect{n}_{\mathrm{x},l}^\mathrm{H}\big]=
\frac{\rho_\mathrm{ad}(1-\rho_\mathrm{ad})}{1-\rho_\mathrm{da}}\diag\bigg(\sum_{i=1}^{K}{\ddot{p}_i}\vect{R}_{il}\bigg)
\\&+(1-\rho_\mathrm{ad})\bigg[\sigma^2+\frac{\rho_\mathrm{ad}}{1-\rho_\mathrm{da}}\sum_{i=1}^{K}\frac{\ddot{p}_i\beta_{il}\kappa_{il}}{\kappa_{il}+1}\bigg]\vect{I}_N,
\end{align}
the diagonal element of the first term in \eqref{Ef_kf_k} can be evaluated as
\begin{equation}
\mathbb{E}[\vect{\hat{h}}_{kl}^\mathrm{H}\vect{n}_{\mathrm{x},l}\vect{n}_{\mathrm{x},l}^\mathrm{H}\vect{\hat{h}}_{kl}]
=\tr\big(\mathbb{E}[\vect{n}_{\mathrm{x},l}\vect{n}_{\mathrm{x},l}^\mathrm{H}]\mathbb{E}[\vect{\hat{h}}_{kl}\vect{\hat{h}}_{kl}^\mathrm{H}]\big)
=d_{kl}.
\end{equation}
According to \eqref{ex-Th1-1}, the second term in \eqref{Ef_kf_k} can be computed as \eqref{ex-hn_yn_yh}.
\begin{figure*}[!t]
\begin{align}\label{ex-hn_yn_yh}
\notag\big[\mathbb{E}[\vect{\hat{h}}_{kl_1}^\mathrm{H}\vect{n}_{\mathrm{y},l_1}\vect{n}_{\mathrm{y},l_2}^\mathrm{H}\vect{\hat{h}}_{kl_2}&]\big]_{|\mathcal{M}_k|}^\mathrm{S}
=\frac{(1-\rho_\mathrm{ad})^2\rho_\mathrm{da}}{1-\rho_\mathrm{da}}\sum_{i=1}^{K}\ddot{p}_i\big[\mathbb{E}[\vect{\hat{h}}_{kl_1}^\mathrm{H}\vect{h}_{il_1}\vect{h}_{il_2}^\mathrm{H}\vect{\hat{h}}_{kl_2}]\big]_{|\mathcal{M}_k|}^\mathrm{S}
\\&=\frac{(1-\rho_\mathrm{ad})^2\rho_\mathrm{da}}{1-\rho_\mathrm{da}}\Bigg\{\sum_{i=1}^{K}\ddot{p}_i\Big[\bm{\lambda}_{k}^i(\bm{\lambda}_{k}^i)^\mathrm{H}
+\big[c_{kl}^i\big]_{|\mathcal{M}_k|}^\Lambda\Big]
+\sum_{i\in\mathcal{P}_k}\ddot{p}_i\Big[\vect{b}_k^i(\vect{b}_k^i)^\mathrm{H}
+\vect{b}_k^i(\bm{\lambda}_{k}^i)^\mathrm{H}+\bm{\lambda}_{k}^i(\vect{b}_k^i)^\mathrm{H}\Big]\Bigg\}
\end{align}
\hrulefill
\end{figure*}
Thus $\mathbb{E}[\vect{f}_k\vect{f}_k^\mathrm{H}]$ can be expressed as in \eqref{ex-Th1-2}.
\begin{figure*}[!t]
\begin{align}\label{ex-Th1-2}
\mathbb{E}[\vect{f}_k\vect{f}_k^\mathrm{H}]=\big[d_{kl}\big]_{|\mathcal{M}_k|}^\Lambda
+\frac{(1-\rho_\mathrm{ad})^2\rho_\mathrm{da}}{1-\rho_\mathrm{da}}\Bigg\{\sum_{i=1}^{K}&\ddot{p}_i\Big[\bm{\lambda}_{k}^i(\bm{\lambda}_{k}^i)^\mathrm{H}
+\big[c_{kl}^i\big]_{|\mathcal{M}_k|}^\Lambda\Big]
+\sum_{i\in\mathcal{P}_k}\ddot{p}_i\Big[\vect{b}_k^i(\vect{b}_k^i)^\mathrm{H}
+\vect{b}_k^i(\bm{\lambda}_{k}^i)^\mathrm{H}+\bm{\lambda}_{k}^i(\vect{b}_k^i)^\mathrm{H}\Big]\Bigg\}
\end{align}
\hrulefill
\end{figure*}
\subsubsection{$\mathbb{E}[\vect{g}_{kk}]\mathbb{E}[\vect{g}_{kk}^\mathrm{H}]$ calculation}
Recalling that
$
\mathbb{E}[\vect{g}_{kk}]=
\big[\mathbb{E}[\vect{\hat{h}}_{kl}^\mathrm{H}\vect{\hat{h}}_{kl}]\big|l\in\mathcal{M}_k\big]^\mathrm{T}
$
where
\begin{equation}
\mathbb{E}[\vect{\hat{h}}_{kl}^\mathrm{H}\vect{\hat{h}}_{kl}]
=\vect{\bar{h}}_{kl}^\mathrm{H}\vect{\bar{h}}_{kl}+(1-\rho_\mathrm{ad})^2\tau\ddot{p}_k\tr(\vect{R}_{kl}\vect{\Psi}_{t_k^\mathrm{p},l}^{-1}\vect{R}_{kl}),
\end{equation}
it can be easily obtained that
$\mathbb{E}[\vect{g}_{kk}]=\bm{\lambda}_{k}^k+\vect{b}_k^k$.
Then, it is clear that
\begin{equation}\label{ex-Th1-3}
\mathbb{E}[\vect{g}_{kk}]\mathbb{E}[\vect{g}_{kk}^\mathrm{H}]=(\bm{\lambda}_{k}^k+\vect{b}_k^k)(\bm{\lambda}_{k}^k+\vect{b}_k^k)^\mathrm{H}.
\end{equation}
By substituting \eqref{ex-Th1-1}, \eqref{ex-Th1-2}, and \eqref{ex-Th1-3} into \eqref{UL-SE-max}, \eqref{UL-SE-MRC} can be obtained and this completes the proof.
\end{appendices}

\bibliography{Ref}
\bibliographystyle{myIEEEtran}

\end{document}